\RequirePackage{amsthm}%

\documentclass[sn-mathphys-num]{sn-jnl}

\usepackage{graphicx}%
\usepackage{multirow}%
\usepackage{amsmath,amssymb,amsfonts}%
\usepackage{mathrsfs}%
\usepackage[title]{appendix}%
\usepackage{xcolor}%
\usepackage{textcomp}%
\usepackage{manyfoot}%
\usepackage{booktabs}%
\usepackage{algorithm}%
\usepackage{algorithmicx}%
\usepackage{algpseudocode}%
\usepackage{listings}%

\theoremstyle{thmstyleone}%
\newtheorem{theorem}{Theorem}
\newtheorem{proposition}[theorem]{Proposition}%

\theoremstyle{thmstyletwo}%
\newtheorem{example}{Example}%

\theoremstyle{thmstylethree}%
\newtheorem{definition}{Definition}%

\usepackage{lipsum}								

\usepackage[utf8]{inputenc}                     
\usepackage[TS1,T1]{fontenc}			        

\usepackage{xpatch}								

\usepackage{booktabs}							

\usepackage{amsmath}							
\usepackage{amssymb}							
\usepackage{amsfonts}							

\usepackage[disable]{todonotes}		            

\usepackage{listings, lstautogobble}			

\usepackage[mathscr]{eucal}						

\usepackage{colortbl}							

\usepackage{amsthm}								

\usepackage{bookmark,hyperref}					

\newcommand\numberstyle[1]{%
	\footnotesize
	\color{SQLcodegray}%
	\ttfamily
	\ifnum#1<10 0\fi#1 |%
}

\makeatletter
\xpatchcmd\algorithmic
{\ALC@linenosize \arabic{ALC@line}\ALC@linenodelimiter}
{\ALC@linenosize \theALC@line\ALC@linenodelimiter}
{}{}

\newcommand\setAlgoLinenoFormat{\renewcommand*{\theALC@line}{\ifnum\value{ALC@line}<10 0\fi\arabic{ALC@line}}}
\makeatother

\let\oldalgorithmic\algorithmic
\renewcommand\algorithmic{\ttfamily\fontseries{l}\selectfont\oldalgorithmic}
\makeatletter
\renewcommand{\ALG@beginalgorithmic}{\footnotesize}
\makeatother

\definecolor{aliceblue}{rgb}{0.94, 0.97, 1.0}
\definecolor{skyblue}{rgb}{0.53, 0.81, 0.92}
\definecolor{brightmaroon}{rgb}{0.95,0.82,0.85}
\definecolor{new}{rgb}{0, 0, 0}

\hypersetup{%
	breaklinks=true,			
	colorlinks=true,			
	urlcolor=blue,				
	citecolor=gray,				
	linkcolor=darkgray,			
	anchorcolor=blue,			
	bookmarksopen=false,		
	linktocpage=true,			
	bookmarksnumbered,			
	pdfstartview={FitH},		
	pdftitle={Skycube émergent},
	pdfauthor={Mickaël Martin Nevot},
	pdfsubject={Article de recherche},
	pdfkeywords = {fouille de données de jeux vidéo, cubes de données fermés, cube de données émergents, Skycube, treillis de concepts, ensembles en Accord},
}

\definecolor{SQLCodeGreen}{rgb}{0,0.6,0}
\definecolor{SQLcodegray}{rgb}{0.5,0.5,0.5}
\definecolor{SQLCodePurple}{HTML}{C42043}
\definecolor{SQLBackgroundcolor}{HTML}{F2F2F2}
\definecolor{SQLBookColor}{cmyk}{0,0,0,0.90}  
\color{SQLBookColor}

\lstdefinestyle{SQLStyle} {
	backgroundcolor=\color{SQLBackgroundcolor},
	commentstyle=\color{SQLCodeGreen},
	keywordstyle=\color{SQLCodePurple},
	numberstyle=\numberstyle,
	stringstyle=\color{SQLCodePurple},
	basicstyle=\footnotesize\ttfamily,
	breakatwhitespace=false,
	breaklines=true,
	captionpos=b,
	keepspaces=true,
	numbers=left,
	numbersep=10pt,
	showspaces=false,
	showstringspaces=false,
	showtabs=false,
	autogobble=true,
	literate = 	{é}{{\'e}}{1}%
	{è}{{\`e}}{1}%
	{à}{{\`a}}{1}%
	{â}{{\^a}}{1}
	{ç}{{\c{c}}}{1}%
	{œ}{{\oe}}{1}%
	{ù}{{\`u}}{1}%
	{É}{{\'E}}{1}%
	{È}{{\`E}}{1}%
	{À}{{\`A}}{1}%
	{Ç}{{\c{C}}}{1}%
	{Œ}{{\OE}}{1}%
	{Ê}{{\^E}}{1}%
	{ê}{{\^e}}{1}%
	{î}{{\^i}}{1}%
	{ï}{{\"i}}{1}
	{ô}{{\^o}}{1}%
	{û}{{\^u}}{1}%
}

\newtheoremstyle{theoremStyle}
{9pt}{9pt}				
{}						
{}						
{\bfseries}{ -}			
{ }						
{}						

\theoremstyle{theoremStyle}

\pgfdeclarelayer{background}
\pgfdeclarelayer{foreground}
\pgfsetlayers{background, main, foreground}

\usetikzlibrary{patterns}
\usetikzlibrary{automata}
\usetikzlibrary{angles}
\usetikzlibrary{quotes}

\ifdefined\proposition\else
\newtheorem{proposition}{Proposition}[section]

\ifdefined\lemma\else
\newtheorem{lemma}[proposition]{Lemme}
\fi

\ifdefined\corollaire\else
\newtheorem{corollaire}[proposition]{Corollaire}
\fi

\ifdefined\theorem\else
\newtheorem{theorem}[proposition]{Théorème}
\fi
\fi

\ifdefined\lemma\else
\newtheorem{lemma}{Lemme}
\fi

\ifdefined\corollaire\else
\newtheorem{corollaire}{Corollaire}
\fi

\ifdefined\theorem\else
\newtheorem{theorem}{Théorème}
\fi

\ifdefined\definition\else
\newtheorem{definition}{Definition}[section]
\fi

\ifdefined\example\else
\newtheorem{example}{Exemple}[section]
\fi

\ifdefined\property\else
\newtheorem{property}{Property}[section]
\fi

\ifdefined\remarques\else
\newtheorem{remarques}{remarques}
\fi

\begin{document}
	
	\title[Ranking Methods for Skyline Queries]{Ranking Methods for Skyline Queries\footnote{This manuscript is currently under review for possible publication in Knowledge and Information Systems (KAIS).}}
	
	
	\author*[1]{\fnm{Mickaël} \sur{Martin Nevot}}\email{mickael.martin-nevot@univ-amu.fr}
	\author*[1]{\fnm{Lotfi} \sur{Lakhal}}\email{lotfi.lakhal@univ-amu.fr}
	
	\affil*[1]{\orgname{Aix-Marseille Université}, \orgaddress{\street{52 AVENUE ESCADRILLE NORMANDIE NIEMEN}, \postcode{13397}, \city{Marseille Cedex 20}, \country{FRANCE}}}
	
	
	\abstract{Multi-criteria decision analysis in databases has been actively studied, especially through the Skyline operator. Yet, few approaches offer a relevant comparison of Pareto optimal, or Skyline, points for high cardinality result sets. We propose to improve the dp-idp method, inspired by tf-idf, a recent approach computing a score for each Skyline point, by introducing the concept of dominance hierarchy. As dp-idp lacks efficiency and does not ensure a distinctive rank, we introduce the RankSky method, the adaptation of Google’s well-known PageRank solution, using a square stochastic matrix, a teleportation matrix, a damping factor, and then a row score eigenvector and the IPL algorithm. For the same reasons as RankSky, and also to offer directly embeddable in DBMS solution, we establish the TOPSIS based CoSky method, derived from both information research and multi-criteria analysis. CoSky automatically ponderates normalized attributes using the Gini index, then computes a score using Salton's cosine toward an ideal point. By coupling multilevel Skyline to dp-idp, RankSky or CoSky, we introduce DeepSky. Implementations of dp-idp, RankSky and CoSky are evaluated experimentally.}

	
	
	
	\maketitle
	
	
	\color{new}
	
	\section{Introduction}
	
	\color{black}
	
	\textcolor{new}{The} Skyline operator (\cite{borzsonySkylineOperator2001}), formerly Pareto set and maximal vectors (\cite{bentleyAverageNumberMaxima1978}), is critical in multi-criteria analysis and has been widely studied. \textcolor{new}{Its main issues with the dataset are, on the one hand, high cardinality and, on the other hand, low correlation.} In these cases, it is often hard to extract significant information since more Skyline points often lead to less differences between them. Efficient ranking overcomes that, but few such approaches have been proposed. 
	
	\textcolor{new}{Some ranking techniques have been proposed,} dp-idp is one of the most recent approaches.
	
	In this paper, we \textcolor{new}{first propose improving} the dp-idp method using dominance hierarchy to perfect score computation of Skyline points, then we define both \textcolor{new}{the RankSky method, which is an adaptation of Google’s well-known PageRank solution and the IPL algorithm, and } the CoSky method (\cite{martinnevotClassementDobjetsSkylines2024})\textcolor{new}{,} to rank efficiently skyline's points ignoring dominance, before presenting DeepSky algorithm, a multilevel Skyline algorithm using CoSky to rank the top-$k$ Skyline points. Finally, implementations of dp-idp and CoSky are presented with experimental evaluation.
	
	
	\section{Use case}
	
	In this paper, we introduce an example applied to Pokémon Showdown!\footnote{Pokémon Showdown! is a popular  Pokémon (Pokémon, possibly because of its Japanese origins, where the plural is not marked on substantives, is, usually, written identically regardless of the plural) open source browser-based online fighting simulator (with millions of monthly visitors).}, and its \texttt{Pokémon} example relation (cf. table~\ref{tab:relation_exemple}).
	
	\begin{table}[htb]
		\caption{\texttt{Pokémon} relation}\label{tab:relation_exemple}
		\centering
		\begin{minipage}{\linewidth}
				\begin{tabular}{c|cc|ccc}
					\toprule
					\texttt{RowId} & \texttt{Player}\footnote{With official Pokémon numbers: n°065: Alakazam, n°080: Slowbro, n°103: Exeggutor, n°113: Chansey, n°121: Starmie, n°128: Tauros, n°143: Snorlax.} & \texttt{Opponent}\footnote{\emph{idem supra}.} & \texttt{Rarity}\footnote{Let $p$ be the percentage of drop for the Pokémon sequence (which is the multiplication of the percentage of drop for each sequence's Pokémon), the \texttt{Rarity} score $r$ is calculated, on a scale of 0 to 10, as such: $\text{if } p = 1 \text{ then } r = 0 \text{, else } r = \lfloor max(\frac{(p - 1) \times 10 - (100 - e \times 0.9)}{e}, 0) \rfloor + 1 $.} & \texttt{Duration}\footnote{In total number of turns in the fight.} & \texttt{Win}\footnote{In percentage.} \\   
					\midrule
					$1$ & $121, 113, 103$ & $121, 113, 121$ & $5$ & $20$ & $70$ \\
					$2$ & $065, 103, 065$ & $065, 143, 065$ & $4$ & $60$ & $50$ \\
					$3$ & $121, 113, 121$ & $065, 103, 065$ & $5$ & $30$ & $60$ \\
					$4$ & $121, 113, 080$ & $065, 143, 065$ & $1$ & $80$ & $60$ \\
					$5$ & $121, 113, 128$ & $121, 113, 121$ & $5$ & $90$ & $40$ \\
					$6$ & $065, 113, 143$ & $065, 113, 143$ & $9$ & $30$ & $50$ \\
					$7$ & $065, 143, 065$ & $121, 113, 143$ & $7$ & $80$ & $40$ \\
					$8$ & $065, 113, 143$ & $065, 103, 065$ & $9$ & $90$ & $30$ \\
					\bottomrule
				\end{tabular}
		\end{minipage}
	\end{table}
	
	\begin{figure}[htb]
		\begin{minipage}{\linewidth}
				\centering
				\includegraphics[width=\linewidth]{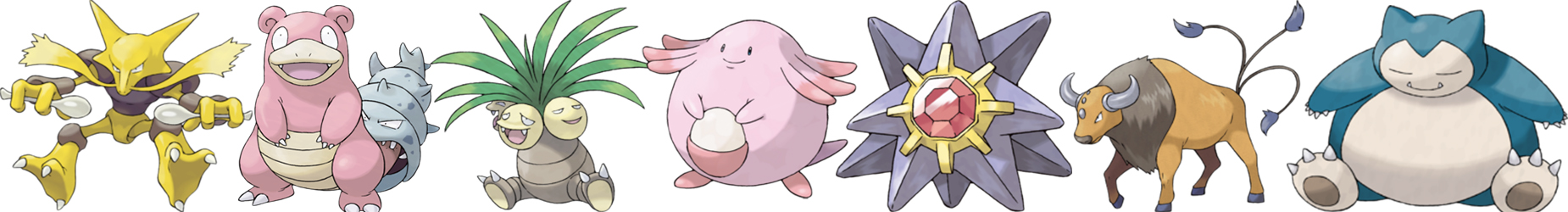}
				\caption[Alakazam, Slowbro, Exeggutor, Chansey, Starmie, Tauros, Snorlax]{The Pokémon of the use case\footnote{From left to right: Alakazam (n°065), Slowbro (n°080), Exeggutor (n°103), Chansey (n°113), Starmie (n°121), Tauros (n°128), Snorlax (n°143); illustration of Pokémon FireRed Version and LeafGreen on Poképédia (\url{https://www.pokepedia.fr/}). }}\label{fig:alakazam_flagadoss_noadkoko_leveinard_staross_tauros-ronflex}
			\end{minipage}
	\end{figure}
	
	\begin{example}
		The \textcolor{new}{the \texttt{Pokémon} example} shown in table~\ref{tab:relation_exemple} is typical of Skyline computation usage. The \texttt{Player} attribute is the player's Pokémon sequence, and \texttt{Opponent}, the opponent's one. Criteria determining the "best Pokémon apparition order in a fight" are the player's Pokémon sequence \texttt{Rarity}, the fight \texttt{Duration} (in rounds) and the player's Pokémon sequence \texttt{Win} rate against the opponent's sequence. For our example, only sequences of three Pokémon are retained, both for the player or the opponent, and to simplify their representation usage we call Pokémon using their number and then substitute a letter for the whole sequence. Thus, the list $121, 113, 128$ corresponds to the three Pokémon sequence Starmie, Chansey and Tauros. For convenience, we round \texttt{Duration} and \texttt{Win} to the nearest five.
	\end{example}
	
	\section{Preliminary Definitions}
	
	\subsection{Preference and dominance}\index{Skyline Preference}\index{Skyline Dominance}
	
	Let $r$ be a relation with attributes $A_1, \dotsc,  A_n$. A Skyline preference on $A_j$ is one of the two following expressions:  $Pref(A_i) = \texttt{MIN}$, or $Pref (A_i) = \texttt{MAX}$. Let $t$ and $t'$ be two tuples of $r$. We consider $t$ dominate $t'$ (noted $t \prec_d t'$) if and only if $t[A_1] \le t'[A_1], \dotsc, t[A_n] \le t'[A_n]$ and $\exists j \in [1..n] : t[A_j] < t'[A_j]$ with:
	\begin{equation}
		(\preceq_d, \prec_d) =
		\left\lbrace
		\begin{array}{l}
			(\le, <) \equiv Pref(A_j) = \texttt{MIN} \\ 
			(\ge, >) \equiv Pref(A_j) = \texttt{MAX}
		\end{array}
		\right.
	\end{equation}
	
	\begin{example}	
		\textcolor{new}{In the \texttt{Pokémon} example shown in table~\ref{tab:relation_exemple}, } \texttt{Rarity} and \texttt{Duration} are criteria to be minimized, while \texttt{Win} is to be maximized \textcolor{new}{(considering that Skyline preferences are mixed). Thus, the tuple of \texttt{RowId} $1$ dominates all others ones considering both \texttt{Duration} and \texttt{Win} criteria (\emph{i.e.}, the sequence $121, 113, 103$ versus the sequence $121, 113, 121$ has the shortest fight, lasting 20 rounds, and this is also the most frequently winning Pokémon sequence with a $70\%$ win rate), whereas the tuple of \texttt{RowId} $4$ dominates all others ones considering only the \texttt{Rarity} criterion (\emph{i.e.}, the sequence containing least rare Pokémon is $121, 113, 080$, regardless of the rarity of the sequence $065, 143, 065$), and the tuple of \texttt{RowId} $2$ does not dominate any other tuple considering any criterion.}
	\end{example}
	
	\subsection{Skyline operator}\index{Skyline operator}
	
	A SQL syntax for the Skyline operator (\cite{borzsonySkylineOperator2001}) has been proposed.
	
	In our example, the SQL query using the Skyline operator is:
	\lstset{style=SQLStyle}
	\begin{lstlisting}[ language=SQL,
						deletekeywords={IDENTITY},
						deletekeywords={[2]INT},
						morekeywords={CLUSTERED, SKYLINE, OF},
						framesep=8pt,
						xleftmargin=40pt,
						framexleftmargin=40pt,
						frame=tb,
						framerule=0pt ]
		SELECT *
		FROM Pokémon
		SKYLINE OF Rarity MIN, Duration MIN, Win MAX
	\end{lstlisting}
	
	And, the associated query without the Skyline operator is as follows:
	\lstset{style=SQLStyle}
	\begin{lstlisting}[ language=SQL,
						deletekeywords={IDENTITY},
						deletekeywords={[2]INT},
						morekeywords={CLUSTERED, SKYLINE, OF},
						framesep=8pt,
						xleftmargin=40pt,
						framexleftmargin=40pt,
						frame=tb,
						framerule=0pt ]
		SELECT *
		FROM Pokémon AS P1
		WHERE NOT EXISTS (
		    SELECT *
		    FROM Pokémon AS P2
		    WHERE (P2.Rarity <= P1.Rarity
		      AND P2.Duration <= P1.Duration
		      AND P2.Win >= P1.Win)
		      AND (P2.Rarity < P1.Rarity
		       OR P2.Duration < P1.Duration
		       OR P2.Win > P1.Win));
	\end{lstlisting}
	
	In a Pareto dominance representation context, we call point belonging to a Skyline, or Skyline point, or Pareto optimal point (belonging to Pareto front), $sp$, each of these tuples, whose set forms a Skyline $S$.
	
	\begin{example}
		In our example, the Skyline is composed of tuples \textcolor{new}{of \texttt{RowId} $1$: $(1, \dotsc, 5, 20, 70)$, $2$: $(2, \dotsc, 4, 60, 50)$ and $4$: $(4, \dotsc, 1, 80, 60)$}. It is the Pokémon sequence set as good as any, or best, regarding all the considered dimension criteria (\texttt{Rarity}, \texttt{Duration} and \texttt{Win}) and best for at least one.
	\end{example}
	
	\section{Skyline ranking}
	
	Many research have been dedicated to Skyline ranking. A metric called Skyline frequency is proposed to order Skyline by prioritizing points with high skyline frequency \textcolor{new}{(\cite{chanHighDimensionalSkylines2006})}. This method works for a large \textcolor{new}{number of dimensions} and experimentation shows a good algorithm efficiency.
	
	Top-$k$ queries (\cite{yiuEfficientProcessingTopk2007}) can be used as an alternative to Skyline queries. An other method is, using user defined regions, to look for regions dominating all the others (\cite{bartoliniFlexibleIntegrationMultimedia2007}). A Skyline ranking approach for a Skycube (\cite{lakhalMultidimensionalSkylineAnalysis2017}) focused on the most informational Skyline points has been proposed (\cite{vlachouRankingSkyDiscovering2010}). This method captures dominance relationships between Skyline points belonging to different subspaces. A new operator has also been introduced to find the most desirable skyline objects, or MDSO (\cite{gaoEfficientAlgorithmsFinding2015}).
	
	\subsection{dp-idp method}\index{Dp-idp}
	
	dp-idp (for dominance power and inverse dominance power) is inspired by the weighting scheme tf-idf (for term frequency-inverse document frequency) use in Information research, attributing to a $t$ term a weight in the $d$ document. The idea is not to determine the occurrence count of each $t$ query term in $d$, but instead the tf-idf weight of each term in $d$. The aim is to find important keywords in a document corpus. In the Skyline context, dominated points have different impact on Skyline points. Thus, their contribution depend on local characteristics corresponding to Skyline points and global characteristics corresponding to the whole Skyline. \textcolor{new}{Thus, dp-idp is build on dominance relation. More specifically, we find the layer of \emph{minima}\footnote{The term "layer of \emph{maxima}" is more common in the literature. Here, however, the term "layer of \emph{minima}" is preferred because small values are assumed to be preferable. It is defined by all minimal points in a given set. It is the first layer, or frontier, of non dominating points in the multidimensional space.} $lm(p, sp)$ where the dominated point $p$ falls in, with respect to $sp$. The dominance of a point is inversely proportional to the number of points that dominate it.} (\cite{valkanasSkylineRankingIR2014}), \emph{i.e.}:
	\begin{equation}\label{eq:dp}
		dp(p, sp) = \frac{1}{lm(p, sp)}
	\end{equation}
	
	\color{new}
	
	\begin{example}
		In the \texttt{Pokémon} example shown in table~\ref{tab:relation_exemple}, for tuples of \texttt{RowId} $1$ and $3$, we have $lm(3, 1) = 2$ and $dp(3, 1) = \frac{1}{lm(3, 1)} = \frac{1}{2}$.
	\end{example}

	\color{black}
	
	dp-idp focuses on dominated points' relative positions to differentiate them by promoting less dominated points: \emph{e.g.} let $sp$ be a Skyline point, if $sp \prec_d p_1, sp \prec_d p_2$ and neither $p_1$ nor $p_2$ dominate each other, they are said similar in regard to $sp$. \textcolor{new}{Whereas}, if $p_1 \prec_d p_2$, then $score(p_1) > score(p_2)$, and consequently $p_1$ has a great significance. A $p \in r \backslash S$ point's $idp$ corresponds to the count of Skyline points dominating $p$. The less frequently a $p$ point appears in a Skyline's dominated points set, the more it is significant:
	\begin{equation}\label{eq:idp}
		idp(p) = \log \frac{|S|}{|\{sp \in S : sp \prec_d p\}|}
	\end{equation}
	
	In order to compute a $p$ dominated point's value $dp$, its relative position regarding a Skyline point $sp$ is essential. Therefore, a given dominated point can participate differently to several Skyline points. Thus, it is mandatory to compute the layer of \emph{minima} $lm(p, sp)$ where $p$, dominated by $sp$, belongs. $p$'s dominance power is then given by the inverse of the layer where it lies. $Score(sp)$ measuring the significance of a Skyline point $sp$ is defined as follows:
	
	\begin{equation}\label{eq:score}
		Score(sp) = \sum_{p : sp \prec_d p} dp(p, sp) \cdot idp(p)
	\end{equation}
	
	A baseline's steps to order a Skyling using dp-idp are:
	\begin{itemize}
		\item computing Skyline points $sp$' layers of \emph{minima};
		\item updating $Score(sp)$ with, for each $p$ point in each layer of \emph{minima} $lm(p, sp)$, the count of points dominating $p$;
		\item ranking Skyline points using these computations.
	\end{itemize}
	
	\textcolor{new}{Unfortunately, this approach is computationally expensive, due to repeated evaluations. It also computes the score of all skyline points, despite any interest in the Top-$k$ results. Finally, it lacks any notion of progressiveness, as the whole Skyline have to be ranked first (\cite{valkanasSkylineRankingIR2014}).}
	
	For these reasons, a more efficient alternative approach, SkyIR, has been proposed (\cite{valkanasSkylineRankingIR2014}). The algorithm is submitted with different priority schemes (Round-robin, pending or Upper Bound). The most efficient configuration is usually the UBS one, named SkyIR-UBS.

	SkyIR-UBS' generation of all layers of minima computations induced complexity is its main issue. We propose to perfect this approach by considering, for each Skyline point, only its closest dominated points, and none of its "indirectly" dominated ones. In this fashion, all $lm(p, sp)$ are composed by a $sp$ and the chain of directly dominated points from $sp$ to $p$.
	
	\subsection{dp-idp improvement}
	
	The dominance relationship can be seen as a hierarchical sort. In other words, a Skyline point has a hierarchical position over its dominated points. This motivated us to correlate the dominance relationship seen earlier with a graph called dominance hierarchy. Using a dominance hierarchy in dp-idp accelerates layers of minima computations, the graph being pruned of all useless edges, and therefore improves its efficiency. Our proposed graph is \textcolor{new}{acyclic} and \textcolor{new}{directed}, portraying a representation of a partially sorted set. An \textcolor{new}{acyclic directed} graph offers a topological order that can be a great representation for a Skyline point $sp$'s hierarchy over its dominated points.
	
	\begin{definition}[Dominance hierarchy]\index{Dominance hierarchy}
		Let $D$ be a points set and $\prec_d$ be a dominance order, therefore the dominance hierarchy (DH) is the ordered set $(D, \prec_d)$'s vertex cover.
	\end{definition}
	
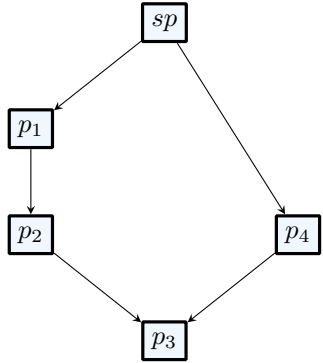
\begin{figure}[htb]
		\centering
			\begin{tikzpicture}[
				line join=bevel,
				box/.style={rectangle, draw=black, fill=aliceblue, very thick, minimum size=5mm}
				]
				
				\node [box] (p3) at (150pt, 0pt) {$p_3$};
				\node [box] (p2) at (100pt, 40pt) {$p_2$};
				\node [box] (p4) at (200pt, 40pt) {$p_4$};
				\node [box] (p1) at (100pt, 80pt) {$p_1$};
				\node [box] (sp) at (150pt, 120pt) {$sp$};
				
				\draw [stealth-] (p3) -- (p2);
				\draw [stealth-] (p3) -- (p4);
				\draw [stealth-] (p2) -- (p1);
				\draw [stealth-] (p4) -- (sp);
				\draw [stealth-] (p1) -- (sp);
			\end{tikzpicture}
		\caption{Dominance hierarchy example}\label{fig:exemple_de_graphe_de_hierarchie_de_dominance}
	\end{figure}

	\begin{example}
		Figure~\ref{fig:exemple_de_graphe_de_hierarchie_de_dominance} portrays a dominance hierarchy with Skyline point $sp$ as its entry node. The dominance order is illustrated by edges between $sp$ and its dominated points ($p_1, p_2, p_3$ and $p_4$).
	\end{example}
	
	We consider the layer of \emph{minima} $lm(p, sp)$ as the minimal path's summits count between $sp$ and $p$ in the dominance hierarchy.
	
	\begin{example}
		To compute the layer of \emph{minima} $lm(p3, sp)$, on the figure~\ref{fig:exemple_de_graphe_de_hierarchie_de_dominance}, we see two paths from $sp$ to $p_3$:
		\begin{enumerate}
			\item First path: $sp \to p_1 \to p_2 \to p_3$.
			\item Second path: $sp \to p_4 \to p_3$.
		\end{enumerate}	
		
		The first path crosses four summits while the second only takes three, therefore the minimal path from $sp$ to $p_3$ is the second and $lm(p_3, sp) = 3$.
	\end{example}
	
	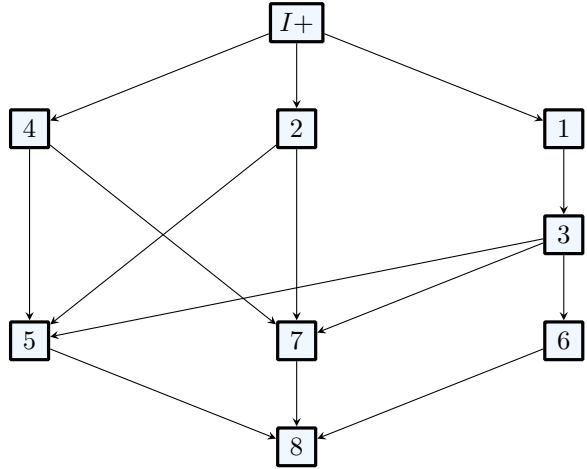
\begin{figure}[htb]
		\centering
			\begin{tikzpicture}[
				line join=bevel,
				box/.style={rectangle, draw=black, fill=aliceblue, very thick, minimum size=5mm}
				]
				
				\node [box] (p01) at (150pt, 0pt) {$8$};
				\node [box] (p11) at (50pt, 40pt) {$5$};
				\node [box] (p12) at (150pt, 40pt) {$7$};
				\node [box] (p13) at (250pt, 40pt) {$6$};
				\node [box] (p21) at (250pt, 80pt) {$3$};
				\node [box] (p31) at (50pt, 120pt) {$4$};
				\node [box] (p32) at (150pt, 120pt) {$2$};
				\node [box] (p33) at (250pt, 120pt) {$1$};
				\node [box] (i) at (150pt, 160pt) {$I+$};
				
				\draw [stealth-] (p01) -- (p11);
				\draw [stealth-] (p01) -- (p12);
				\draw [stealth-] (p01) -- (p13);
				\draw [stealth-] (p11) -- (p31);
				\draw [stealth-] (p11) -- (p32);
				\draw [stealth-] (p11) -- (p21);
				\draw [stealth-] (p12) -- (p31);
				\draw [stealth-] (p12) -- (p32);
				\draw [stealth-] (p12) -- (p21);
				\draw [stealth-] (p13) -- (p21);
				\draw [stealth-] (p21) -- (p33);
				\draw [stealth-] (p31) -- (i);
				\draw [stealth-] (p32) -- (i);
				\draw [stealth-] (p33) -- (i);
			\end{tikzpicture}
		\caption{Example's dominance hierarchy graph}
		\label{fig:graphe_de_hierarchie_de_dominance_de_pokemon_showdown}
	\end{figure}
	
	\begin{example}
		Considering again the \texttt{Pokémon} relation (cf. table~\ref{tab:relation_exemple}), the dominance hierarchy graph portraying dominance relationships between Skyline points and dominated points is given by figure~\ref{fig:graphe_de_hierarchie_de_dominance_de_pokemon_showdown} ($I^+$ being the ideal theoretical point, or abstract, dominating all Skyline points, and points are represented by their \texttt{RowId}).
		
		To compute each Skyline point's score, we use the formula~\ref{eq:score}. Thus, Skyline points' sorting in descending order is either $1, 2, 4$, or $1, 4, 2$. \textcolor{new}{Intuitively, without taking formula~\ref{eq:score} into account, one can see from figure~\ref{fig:graphe_de_hierarchie_de_dominance_de_pokemon_showdown} that from point $1$, we can reach five points via some path, while from points $2$ and $4$, we can reach only three points in both cases. This is why point $1$ is more important than points $2$ and $4$ and why points $2$ and $4$ are equally important.}
		
		In computation formula of idp, each point dominated by every Skyline points will have a $0$ score. These points do not alter the Skyline ranking as they contribute equally to all Skyline points (\cite{valkanasSkylineRankingIR2014}).
		
		Based on our results (cf. table~\ref{tab:calcul_de_score_avec_la_methode_dp_idp_amelioree}), we show the failure of dp-idp method to always distinguish two Skyline points as with values of \texttt{RowId} $2$ and $4$.
		
		\begin{table}[htb]
			\caption{Improved dp-idp method score computation}\label{tab:calcul_de_score_avec_la_methode_dp_idp_amelioree}
			\centering
			\begin{minipage}{\linewidth}
				\centering
				\begin{tabular}{c|c|c|c}
					\toprule
					\texttt{RowId} of $sp$ & \texttt{RowId} of $p$ & $lm(p, sp)$ & $Score(sp)$ \\
					\midrule
					$1$ & $3$ & $lm(3, 1) = 2$ & $0.398$ \\
					& $5$ & $lm(5, 1) = 3$ &         \\
					& $6$ & $lm(6, 1) = 3$ &         \\
					& $7$ & $lm(7, 1) = 3$ &         \\
					& $8$ & $lm(8, 1) = 4$ &         \\
					\midrule
					$2$ & $5$ & $lm(5, 2) = 2$ & $0$     \\
					& $7$ & $lm(7, 2) = 2$ &         \\
					& $8$ & $lm(8, 2) = 3$ &         \\
					\midrule
					$4$ & $5$ & $lm(5, 4) = 2$ & $0$     \\
					& $7$ & $lm(7, 4) = 2$ &         \\
					& $8$ & $lm(8, 4) = 3$ &         \\
					\bottomrule
				\end{tabular}
			\end{minipage}
		\end{table}
	\end{example}
	
	\color{new}
	
	\subsubsection{Improved algorithm}
	
	\color{black}
	
	dp-idp algorithm with dominance hierarchy is composed of four parts: dominanceMatrix, coverageGraph, lm and score$_{dp-idp}$. it is described in algorithm~\ref{algo:dp-idp}.
	
	\begin{algorithm}[htbp]
		\caption{dp-idp with dominance hierarchy\label{algo:dp-idp}}
		\begin{algorithmic}
			\Require~ \\
			$r$ relation. \\
			\Ensure~ \\
			The dp-idp score table $score$. \\
			\State
			\Comment{Calling four algorithm parts}
			\State $m_{\prec_d}, S_{\setminus \mathcal{D}+lm} := \texttt{dominanceMatrix}(r)$;
			\State $m_{\prec_d}, S_{\setminus \mathcal{D}+lm}, S_{\prec_d}C, idpC := $
			\State $\texttt{coverageGraph}(m_{\prec_d}, S_{\setminus \mathcal{D}+lm})$;
			\State $S_{\setminus \mathcal{D}+lm} := \texttt{lm}(m_{\prec_d}, S_{\setminus \mathcal{D}+lm}, S_{\prec_d}C)$;
			\Return $\texttt{score}_{dp-idp}(S_{\setminus \mathcal{D}+lm}, idpC)$;
		\end{algorithmic}
	\end{algorithm}
	
	\begin{example}
		The table~\ref{tab:relation_exemple_2} portrays the \texttt{Pokémon} relation which, for commodity reasons, carries no attribute or details, only tuples' \texttt{RowId}.
		
		\begin{table}[htb]
			\caption{The simplified $\texttt{Pokémon}_{1}$ relation}\label{tab:relation_exemple_2}
			\centering
			\begin{minipage}{\linewidth}
				\centering
				\begin{tabular}{c|ccc}
					\toprule
					\texttt{RowId} & \texttt{Rarity} & \texttt{Duration} & \texttt{Win} \\
					\midrule
					$1$ & $5$ & $20$ & $70$ \\
					$2$ & $4$ & $60$ & $50$ \\
					$3$ & $5$ & $30$ & $60$ \\
					$4$ & $1$ & $80$ & $60$ \\
					$5$ & $5$ & $90$ & $40$ \\
					$6$ & $9$ & $30$ & $50$ \\
					$7$ & $7$ & $80$ & $40$ \\
					$8$ & $9$ & $90$ & $30$ \\
					\bottomrule
				\end{tabular}
			\end{minipage}
		\end{table}
	\end{example}
	
	\begin{algorithm}[htbp]
		\caption{dominanceMatrix ($\mathcal{O}(|r|^2 \cdot |\mathcal{D}|)$)\label{algo:matrice_des_dominants}}
		\begin{algorithmic}
			\Require~ \\
			$r$ relation. \\
			\Ensure~ \\
			The dominance square matrix $m_{\prec_d}$. \\
			The dimensionless Skyline intended to receive $lm(sp, p)$ of dominated points $S_{\setminus \mathcal{D}+lm}$. \\
			\State $\mathcal{D} := \{d_1, \dotsc, d_n \}$: $r$'s dimensions set.
			\State $m_{\prec_d}$: an array of $|r|$ arrays of $|r|$ false values
			\State $idpC$: a dimensionless relation of $|r|$ tuples with same $\texttt{RowId}$ as $r$
			\For{$i := 0, \dots, |r| - 1$}
			\For{$j := 0, \dots, |r| - 1$}
			\If{$i \neq j$}
			\State $t_i$: $t_i \in r, t_i[\texttt{RowId}] = i$;
			\State $t_j$: $t_j \in r, t_j[\texttt{RowId}] = j$;
			\State $gre := true$;
			\ForAll{$d_k \in \mathcal{D}$}
			\State
			\Comment{Dominance computation}
			\State
			\Comment{Comparator $>$ if $\forall \{d_1, \dotsc, d_n \} \in \mathcal{D}, Pref(d_k) = MIN$}
			\If{$t_j[d_k] > t_i[d_k]$}
			\State $gre := false$;
			\Comment{$t_j \nprec_d t_i$}
			\State \textbf{break}
			\EndIf
			\EndFor
			\If{$gre$}
			\State $t'_i$: $t'_i \in S_{\setminus \mathcal{D}}, t'_i[\texttt{RowId}] = i$;
			\State
			\Comment{Adding the edge to the graph}
			\State $m_{\prec_d}[i][j] := true$;
			\Comment{$t_j \prec_d t_i$}
			\State $S_{\setminus \mathcal{D}} := S_{\setminus \mathcal{D}} \setminus t'_i$;
			\Comment {$t_i \notin S$}
			\EndIf
			\EndIf
			\EndFor
			\EndFor
			\State $S_{\setminus \mathcal{D}+lm}$: a relation of $|r|$ dimensions of $S_{\setminus \mathcal{D}+lm}$ tuples $(0, \dots, 0)$ with same $\texttt{RowId}$ as $S_{\setminus \mathcal{D}+lm}$
			\Return $m_{\prec_d}, S_{\setminus \mathcal{D}+lm}$;
		\end{algorithmic}
	\end{algorithm}
	
	\begin{example}
		With the \texttt{Pokémon} relation (cf. table~\ref{tab:relation_exemple_2}), result of the algorithm~\ref{algo:matrice_des_dominants}, the table~\ref{tab:matrice_des_dominants} portrays the dominance matrix $m_{\prec_d}$, in a table representation with \texttt{RowId} both in row and column, and the table~\ref{tab:skyline_sans_dimension_initial} portrays the dimensionless Skyline\footnote{\textcolor{new}{We call dimensionless Skyline a Skyline without dimensions other than \texttt{RowId}.}} intended to receive layers of \emph{minima} $S_{\setminus \mathcal{D}+lm}$. \textcolor{new}{However, note that the matrix $m_{\prec_d}$ requires a large amount of memory.} With the table~\ref{tab:matrice_des_dominants}, the tuple having \texttt{RowId} $6$ is dominated both by \texttt{RowId} $1$ and by \texttt{RowId} $3$. The table~\ref{tab:skyline_sans_dimension_initial} presents Skyline tuples' \texttt{RowId} on rows, and relation tuples' \texttt{RowId} on columns.
		
		\begin{table}[htb]
			\caption{Dominance matrix $m_{\prec_d}$}\label{tab:matrice_des_dominants}
			\centering
			\begin{minipage}{\linewidth}
				\centering
				\begin{tabular}{c|c|c|c|c|c|c|c|c}
					\toprule
					\texttt{RowId} & \texttt{$1$} & \texttt{$2$} & \texttt{$3$} & \texttt{$4$} & \texttt{$5$} & \texttt{$6$} & \texttt{$7$} & \texttt{$8$} \\
					\midrule
					$1$ &  &  &  &  &  &  &  &  \\
					\midrule
					$2$ &  &  &  &  &  &  &  &  \\
					\midrule
					$3$ & \checkmark &  &  &  &  &  &  &  \\
					\midrule
					$4$ &  &  &  &  &  &  &  &  \\
					\midrule
					$5$ & \checkmark & \checkmark & \checkmark & \checkmark &  &  &  &  \\
					\midrule
					$6$ & \checkmark &  & \checkmark &  &  &  &  &  \\
					\midrule
					$7$ & \checkmark & \checkmark & \checkmark & \checkmark &  &  &  &  \\
					\midrule
					$8$ & \checkmark & \checkmark & \checkmark & \checkmark & \checkmark & \checkmark & \checkmark &  \\
					\bottomrule
				\end{tabular}
			\end{minipage}
		\end{table}
		
		\begin{table}[htb]
			\caption{Dimensionless Skyline intended to receive \emph{lm} $S_{\setminus \mathcal{D}+lm}$}\label{tab:skyline_sans_dimension_initial}
			\centering
			\begin{minipage}{\linewidth}
				\centering
				\begin{tabular}{c|cccccccc}
					\toprule
					\texttt{RowId} & \texttt{$1$} & \texttt{$2$} & \texttt{$3$} & \texttt{$4$} & \texttt{$5$} & \texttt{$6$} & \texttt{$7$} & \texttt{$8$} \\
					\midrule
					$1$ & 0 & 0 & 0 & 0 & 0 & 0 & 0 & 0 \\
					$2$ & 0 & 0 & 0 & 0 & 0 & 0 & 0 & 0 \\
					$4$ & 0 & 0 & 0 & 0 & 0 & 0 & 0 & 0 \\
					\bottomrule
				\end{tabular}
			\end{minipage}
		\end{table}
	\end{example}
	
	The algorithm~\ref{algo:graphe_de_couverture}, coverageGraph, updates the dominance matrix $m_{\prec_d}$ considering only the coverage graph. The dimensionless Skyline  $S_{\setminus \mathcal{D}+lm}$ is also updated to receive layers of \emph{minima} to compute, indicated by $1$. \textcolor{new}{The purpose of the algorithm~\ref{algo:graphe_de_couverture} is to remove all transitive dominance relationships from the dominance matrix $m_{\prec_d}$. It is a transitive reduction, not a transitive closure.)} To optimize, Skyline points' dominance cardinalities $S_{\prec_d}C$ and cardinalities $idpC$ needed to compute idp are also retrieved during the coverage graph navigation.
	
	\begin{algorithm}[htbp]
		\caption{coverageGraph ($\mathcal{O}(|r|^3/2)$)\label{algo:graphe_de_couverture}}
		\begin{algorithmic}
			\Require~ \\
			The dominance square matrix $m_{\prec_d}$. \\
			The dimensionless Skyline intended to receive $lm(sp, p)$ of dominated points $S_{\setminus \mathcal{D}+lm}$. \\
			\Ensure~ \\
			The dominance square matrix $m_{\prec_d}$. \\
			The dimensionless Skyline intended to receive $lm(sp, p)$ of dominated points $S_{\setminus \mathcal{D}+lm}$. \\
			$S$ points' dominance cardinalities $S_{\prec_d}C$. \\
			$idpC$ cardinalities of $sp \in S, sp \prec_d p$. \\
			\State $S_{\prec_d}C$: an array of $|S_{\setminus \mathcal{D}+lm}|$ 0
			\State $idpC$: an array of $|m_{\prec_d}|$ 0
			\For{$i := 0, \dots, |m_{\prec_d}| - 1$}
			\If{$t_i, t[\texttt{RowId}] = i, t_i \notin S_{\setminus \mathcal{D}+lm}$}
			\For{$j := 0, \dots, |m_{\prec_d}| - 1$}
			\If{$m_{\prec_d}[i][j]$}
			\If{$t_j, t[\texttt{RowId}] = j, t_j \in S_{\setminus \mathcal{D}+lm}$}
			\State
			\Comment{Marking dominances}
			\State $t_j[i] := 1$
			\State
			\Comment{Updating dominance cardinalities}
			\State $S_{\prec_d}C[j] := S_{\prec_d}C[j] + 1$;
			\State $idpC[i] := idpC[i] + 1$;
			\Else
			\For{$k = 0, \dots, |m_{\prec_d}| - 1$}
			\If{$m_{\prec_d}[j][k]$}
			\State
			\Comment{Deleting the useless coverage graph edge}
			\State $m_{\prec_d}[j][k] := false$;
			\EndIf
			\EndFor
			\EndIf
			\EndIf
			\EndFor
			\EndIf
			\EndFor
			\Return $m_{\prec_d}, S_{\setminus \mathcal{D}+lm}, S_{\prec_d}C, idpC$;
		\end{algorithmic}
	\end{algorithm}
	
	\color{new}
	
	\begin{example}
		Considering again the \texttt{Pokémon} relation (cf. table~\ref{tab:relation_exemple_2}), with the algorithm~\ref{algo:graphe_de_couverture} we have $m_{\prec_d}[5][1] := false$ because of $m_{\prec_d}[3][1] = true$ and $m_{\prec_d}[5][3] = true$.
	\end{example}

	\color{black}
	
	\begin{example}
		With the \texttt{Pokémon} relation (cf. table~\ref{tab:relation_exemple_2}), result of the algorithm~\ref{algo:matrice_des_dominants}, the table~\ref{tab:matrice_des_dominants_du_graphe_de_couverture} portrays the dominance matrix $m_{\prec_d}$ based on the coverage graph and the table~\ref{tab:skyline_sans_dimension_pret_pour_lm} portrays the dimensionless Skyline intended to receive layers of \emph{minima} $S_{\setminus \mathcal{D}+lm}$, merged with, on column, Skyline points' dominance cardinalities $S_{\prec_d}C$ and, on row, cardinalities $idpC$ needed to compute idp. We adopt this representation for more practicality and comprehension. With the table~\ref{tab:matrice_des_dominants_du_graphe_de_couverture}, we have the dominance hierarchy graph for Pokémon Showdown! (cf. figure~\ref{fig:graphe_de_hierarchie_de_dominance_de_pokemon_showdown}). And, with the table~\ref{tab:skyline_sans_dimension_pret_pour_lm}, we see (with values $1$) all dominances of each Skyline point (whose \texttt{RowId} are on rows). We also see the count of dominated points for each Skyline point in the column $S_{\prec_d}C$, and the count of dominant Skyline points for each relation point (whose \texttt{RowId} are on columns) with the row $idpC$.
		
		\begin{table}[htb]
			\caption{$m_{\prec_d}$ based on coverage graph}\label{tab:matrice_des_dominants_du_graphe_de_couverture}
			\centering
			\begin{minipage}{\linewidth}
				\centering
				\begin{tabular}{c|c|c|c|c|c|c|c|c}
					\toprule
					\texttt{RowId} & \texttt{$1$} & \texttt{$2$} & \texttt{$3$} & \texttt{$4$} & \texttt{$5$} & \texttt{$6$} & \texttt{$7$} & \texttt{$8$} \\
					\midrule
					$1$ &  &  &  &  &  &  &  &  \\
					\midrule
					$2$ &  &  &  &  &  &  &  &  \\
					\midrule
					$3$ & \checkmark &  &  &  &  &  &  &  \\
					\midrule
					$4$ &  &  &  &  &  &  &  &  \\
					\midrule
					$5$ &  & \checkmark & \checkmark & \checkmark &  &  &  &  \\
					\midrule
					$6$ &  &  & \checkmark &  &  &  &  &  \\
					\midrule
					$7$ &  & \checkmark & \checkmark & \checkmark &  &  &  &  \\
					\midrule
					$8$ &  &  &  &  & \checkmark & \checkmark & \checkmark &  \\
					\bottomrule
				\end{tabular}
			\end{minipage}
		\end{table}
		
		\begin{table}[htb]
			\caption{$S_{\setminus \mathcal{D}+lm}$ intended to receive \emph{lm} with $S_{\prec_d}C$ and $idpC$ }\label{tab:skyline_sans_dimension_pret_pour_lm}
			\centering
			\begin{minipage}{\linewidth}
				\centering
				\begin{tabular}{c|cccccccc|c}
					\toprule
					\texttt{RowId} & \texttt{$1$} & \texttt{$2$} & \texttt{$3$} & \texttt{$4$} & \texttt{$5$} & \texttt{$6$} & \texttt{$7$} & \texttt{$8$} & $S_{\prec_d}C$ \\
					\midrule
					$1$    & 0 & 0 & 1 & 0 & 1 & 1 & 1 & 1 & 5 \\
					$2$    & 0 & 0 & 0 & 0 & 1 & 0 & 1 & 1 & 3 \\
					$4$    & 0 & 0 & 0 & 0 & 1 & 0 & 1 & 1 & 3 \\
					\midrule
					\multicolumn{1}{c|}{$idpC$} & 
					\multicolumn{1}{c|}{0} & 
					\multicolumn{1}{c|}{0} & 
					\multicolumn{1}{c|}{1} & 
					\multicolumn{1}{c|}{0} & 
					\multicolumn{1}{c|}{3} & 
					\multicolumn{1}{c|}{1} & 
					\multicolumn{1}{c|}{3} & 
					\multicolumn{1}{c|}{3} & / \\
					\bottomrule
				\end{tabular}
			\end{minipage}
		\end{table}
	\end{example}
	
	The algorithm~\ref{algo:lm}, lm, computes layers of \emph{minima} for each Skyline point based on the coverage graph from the dominance matrix $m_{\prec_d}$. Results are stored in the dimensionless Skyline $S_{\setminus \mathcal{D}+lm}$. As suited as a recursive function might seem for this task, it would be less efficient.
	
	\begin{algorithm}[htbp]
		\caption{lm ($\mathcal{O}(|r| \cdot |S|)$)\label{algo:lm}}
		\begin{algorithmic}
			\Require~ \\
			The dominance square matrix $m_{\prec_d}$. \\
			The dimensionless Skyline intended to receive $lm(sp, p)$ of dominated points $S_{\setminus \mathcal{D}+lm}$. \\
			$S$ points' dominance cardinalities $S_{\prec_d}C$. \\
			\Ensure~ \\
			The $S$ Skyline with layers $lm(sp, p)$ of dominated points, $S_{lm}$. \\
			\ForAll{$t \in S_{\setminus \mathcal{D}+lm}$}
			\State $i := t[\texttt{RowId}]$;
			\State $layer := i$
			\State $depth := 2$
			\While{$S_{\prec_d}C[i] > 0$}
			\State $layer_{+1} := layer$
			\For{$j = 0, \dots, |m_{\prec_d}| - 1$}
			\If{$m_{\prec_d}[j][layer] \wedge t[j] > 0$}
			\If{$t[j] = 1 \vee t[j] \neq 1 \wedge t[j] > depth$}
			\State
			\Comment{Adjusting $lm(t, t_j), t_j \in r$ depth}
			\State $t[j] := t[j] + depth$;
			\EndIf
			\State $S_{\prec_d}C[i] := S_{\prec_d}C[i] - 1$;
			\Comment{Dominance handled}
			\State $layer_{+1} := j$
			\If{$S_{\prec_d}C[i] = 0$}
			\State
			\Comment{Processing of Skyline point stops when all dominances are handled}
			\State \textbf{break}
			\EndIf
			\EndIf
			\State $depth := depth + 1$
			\If{$layer_{+1} = layer$}
			\State \textbf{break}
			\Comment{No more layer to handle}
			\EndIf
			\State $layer := layer_{+1}$
			\EndFor
			\EndWhile
			\EndFor
			\Return{$S_{\setminus \mathcal{D}+lm}$};
		\end{algorithmic}
	\end{algorithm}
	
	\begin{example}
		With the \texttt{Pokémon} relation (cf. table~\ref{tab:relation_exemple_2}), result of the algorithm~\ref{algo:lm}, the table~\ref{tab:skyline_sans_dimension_final} portrays the dimensionless Skyline with layers of \emph{minima} $S_{\setminus \mathcal{D}+lm}$. We find, for example, $lm(3, 1) = 2$, $lm(8, 1) = 4$ or $lm(2, 5) = 2$.
		
		\begin{table}[htb]
			\caption{Dimensionless Skyline with \emph{lm} $S_{\setminus \mathcal{D}+lm}$}\label{tab:skyline_sans_dimension_final}
			\centering
			\begin{minipage}{\linewidth}
				\centering
				\begin{tabular}{c|cccccccc}
					\toprule
					\texttt{RowId} & \texttt{$1$} & \texttt{$2$} & \texttt{$3$} & \texttt{$4$} & \texttt{$5$} & \texttt{$6$} & \texttt{$7$} & \texttt{$8$} \\
					\midrule
					$1$ & 0 & 0 & 2 & 0 & 3 & 3 & 3 & 4 \\
					$2$ & 0 & 0 & 0 & 0 & 2 & 0 & 2 & 3 \\
					$4$ & 0 & 0 & 0 & 0 & 2 & 0 & 2 & 3 \\
					\bottomrule
				\end{tabular}
			\end{minipage}
		\end{table}
	\end{example}
	
	The algorithm~\ref{algo:score}, score$_{dp-idp}$, computes Skyline points scores using the dp-idp method. To this end, it requires layers of \emph{minima} stored in the dimensionless Skyline $S_{\setminus \mathcal{D}+lm}$, useful for dp computation, and also, for each relation point, the count $idpC$ of Skyline points dominating it, useful for idp computation.
	
	\begin{algorithm}[htbp]
		\caption{score$_{dp-idp}$ ($\mathcal{O}(|S| \cdot |\mathcal{D}|)$)\label{algo:score}}
		\begin{algorithmic}
			\Require~ \\
			The dimensionless Skyline with $lm(sp, p)$ of dominated points $S_{\setminus \mathcal{D}+lm}$. \\
			$S$ points' dominance cardinalities $S_{\prec_d}C$. \\
			\Ensure~ \\
			The array of dp-idp scores, $score$. \\
			\State $score$: an associative array of $|S_{\setminus \mathcal{D}+lm}|$ 0
			\ForAll{$t \in S_{\setminus \mathcal{D}+lm}$}
			\State $i := t[\texttt{RowId}]$;
			\For{$j = 0, \dots, |t| - 1$}
			\If{$t[j] > 0$}
			\State $score[i] := score[i] + 1 / t[j] \times \log(|S_{\setminus \mathcal{D}+lm}| / idpC[j])$;
			\State
			\Comment{cf. formula~\ref{eq:score}}
			\EndIf
			\EndFor
			\EndFor
			\Return $score$;
		\end{algorithmic}
	\end{algorithm}
	
	\begin{example}
		With the \texttt{Pokémon} relation (cf. table~\ref{tab:relation_exemple_2}), result of the algorithm~\ref{algo:score}, the table~\ref{tab:tableau_de_score_de_dp_idp_ameliore} portrays respective scores of Skyline tuples' \texttt{RowId} $1$, $2$ and $4$.
		
		\begin{table}[htbp]
			\caption{Improved dp-idp score table}\label{tab:tableau_de_score_de_dp_idp_ameliore}
			\centering
			\begin{minipage}{\linewidth}
				\centering
				\begin{tabular}{c|ccc}
					\toprule
					\texttt{RowId} & \texttt{$1$} & \texttt{$2$} & \texttt{$4$} \\
					\midrule
					$score$ & 0.398 & 0 & 0 \\
					\bottomrule
				\end{tabular}
			\end{minipage}
		\end{table}
	\end{example}
	
	\subsection{Preparation and comments for methods on vectorial model}\label{ssec:etape_0_preparation_et_remarques_prealables}
	
	\subsubsection{Convertion of Skyline preferences}\label{sssec:conversion_d_une_preference_skyline}
	
	Let $r$ be a relation composed of attributes $A_1, \dotsc,  A_m$. The conversion from $Pref(A_i) = MIN$ to $Pref(A_i) = MAX$, or its reciprocal, is difficult in ordering schemas based on vectorial model like RankSky with matrix calculation or CoSky with Salton cosine. Inversion of Skyline attribute's values is therefore preferable to complement operation (and to a conversion from minimum to maximum or from maximum to minimum).
	
	Typically, complement values computation formula used is $\forall t \in r, t[A_i'] = \bigvee A_i - t[A_i]$ where $\bigvee A_i$ is the \emph{supremum} of $A_i$, a theoretical value high enough to ensure each converted value remains positive and interpretable. $\bigvee A_i$ can be either the $A_i$ attribute's current maximal value (\emph{optimum}) or its highest possible value (maximum). An alternative approach will be to use $\forall t \in r, t[A_i'] = \bigwedge A_i + \bigvee A_i - t[A_i]$ where $\bigwedge A_i$ is  $A_i$'s \emph{infimum}. With this approach, \emph{infimum} and \emph{supremum} are, most of the time, respectively the minimum and maximum.
	
	Value inversion is less problematic, mainly for strictly positive values. This method is indifferent to bounds' values, thus computing the \emph{supremum} is irrelevant, and proportions are preserved. 
	
	However, even then this is not a perfect conversion due to data dispersion (range, variance, mean absolute deviation, sum, etc.).
	
	\begin{example}
		With \texttt{Pokémon} relation (cf. table~\ref{tab:relation_exemple}), to convert the highest possible \texttt{Win} rate to the lowest \texttt{Win}$^{-1}$ rate (or \texttt{Loss}), we can invert corresponding values. This produce table~\ref{tab:relation_exemple_4}, where, for convenience reasons, no attribute or comment are left, only tuples' \texttt{RowId}. However, even though dominance relations between tuples are preserved after transformation, Skyline points attributes values sums (for \texttt{RowId} $1$, $2$ and $4$), respectively $\Sigma = 160$ and $\Sigma = 9 / 140 \approx 0,0643$ are different. In this case, most normalizations, such as CoSky sum one, give different results when $Pref(\texttt{Win}) = MAX$ or $Pref(\texttt{Win}^{-1}) = MIN$.
	\end{example}
	
	\subsubsection{Unification of Skyline preferences}\label{sssec:unification_des_preferences_skyline}
	
	In vector space model context as with matrix calculation, like in RankSky, or Salton cosine, like in CoSky,, we must have comparable values, and therefore identical Skyline preferences. Indeed, in the case of vectors or matrix calculation, all preferences must be set to $\texttt{MAX}$. Also, for Salton cosine to give a significative similarity measure, it is mandatory for vectors' magnitudes to be on comparable scales and to have similar units.
	
	This way, we can compute either a "mimimum ideal" score or a "maximum ideal" score, even though we recommend, for best precision and quickest computation to unify Skyline preferences selecting the most frequent one, except when a specific unification is required as with PageRank.
	
	Skyline preferences unification process must be done prior to any computation.
	
	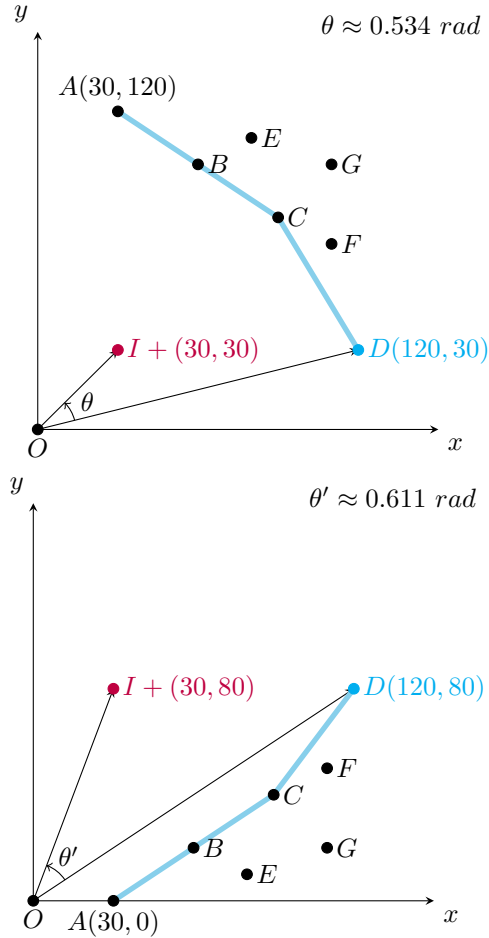
\begin{figure}[htbp]
		\begin{minipage}{\linewidth}
			\centering
				\begin{tikzpicture}[
					line join=bevel
					]
					
					\draw [-stealth] (0pt, 0pt) -- (150pt, 0pt) node[anchor=north west] {$x$};
					\draw [-stealth] (0pt, 0pt) -- (0pt, 150pt) node[anchor=south east] {$y$};
					
					\draw [-stealth] (0pt, 0pt) -- (30pt, 30pt);
					\draw [-stealth] (0pt, 0pt) -- (120pt, 30pt);
					
					\draw[skyblue, line width=2pt] (30pt, 120pt) -- (60pt, 100pt) -- (90pt, 80pt) -- (120pt, 30pt);
					
					\filldraw[purple] (30pt, 30pt) circle (2pt) node[anchor=west]{$I+ (30, 30)$};
					\filldraw[black] (0pt, 0pt) circle (2pt) node[anchor=north]{$O$};
					
					\filldraw[black] (30pt, 120pt) circle (2pt) node[anchor=south]{$A (30, 120)$};
					\filldraw[black] (60pt, 100pt) circle (2pt) node[anchor=west]{$B$};
					\filldraw[black] (90pt, 80pt) circle (2pt) node[anchor=west]{$C$};
					\filldraw[cyan] (120pt, 30pt) circle (2pt) node[anchor=west]{$D (120, 30)$};
					
					\filldraw[black] (80pt, 110pt) circle (2pt) node[anchor=west]{$E$};
					\filldraw[black] (110pt, 70pt) circle (2pt) node[anchor=west]{$F$};
					\filldraw[black] (110pt, 100pt) circle (2pt) node[anchor=west]{$G$};
					
					\coordinate (D) at (120pt, 30pt);
					\coordinate (O) at (0pt, 0pt);
					\coordinate (I) at (30pt, 30pt);
					\pic [draw, ->, "$\theta$", angle eccentricity=1.5] {angle = D--O--I};
					
					\matrix [below left] at (current bounding box.north east) {
						\node [label=right:$\theta \approx 0.534~rad$] {}; \\
					};
				\end{tikzpicture}
				\begin{tikzpicture}[
					line join=bevel
					]
					
					\draw [-stealth] (0pt, 0pt) -- (150pt, 0pt) node[anchor=north west] {$x$};
					\draw [-stealth] (0pt, 0pt) -- (0pt, 150pt) node[anchor=south east] {$y$};
					
					\draw [-stealth] (0pt, 0pt) -- (30pt, 80pt);
					\draw [-stealth] (0pt, 0pt) -- (120pt, 80pt);
					
					\draw[skyblue, line width=2pt] (30pt, 0pt) -- (60pt, 20pt) -- (90pt, 40pt) -- (120pt, 80pt);
					
					\filldraw[purple] (30pt, 80pt) circle (2pt) node[anchor=west]{$I+ (30, 80)$};
					\filldraw[black] (0pt, 0pt) circle (2pt) node[anchor=north]{$O$};
					
					\filldraw[black] (30pt, 0pt) circle (2pt) node[anchor=north]{$A (30, 0)$};
					\filldraw[black] (60pt, 20pt) circle (2pt) node[anchor=west]{$B$};
					\filldraw[black] (90pt, 40pt) circle (2pt) node[anchor=west]{$C$};
					\filldraw[cyan] (120pt, 80pt) circle (2pt) node[anchor=west]{$D (120, 80)$};
					
					\filldraw[black] (80pt, 10pt) circle (2pt) node[anchor=west]{$E$};
					\filldraw[black] (110pt, 50pt) circle (2pt) node[anchor=west]{$F$};
					\filldraw[black] (110pt, 20pt) circle (2pt) node[anchor=west]{$G$};
					
					\coordinate (D) at (120pt, 80pt);
					\coordinate (O) at (0pt, 0pt);
					\coordinate (I) at (30pt, 80pt);
					\pic [draw, ->, "$\theta'$", angle eccentricity=1.5] {angle = D--O--I};
					
					\matrix [below left] at (current bounding box.north east) {
						\node [label=right:$\theta' \approx 0.611~rad$] {}; \\
					};
				\end{tikzpicture}
			\caption{Pref. $(\texttt{MIN}, \texttt{MIN})$ / $(\texttt{MIN}, \texttt{MAX})$}\label{fig:cosinus_de_salton_avec_preferences_skyline_unifiees_min_min}
		\end{minipage}
	\end{figure}
	
	\begin{example}
		In figure~\ref{fig:cosinus_de_salton_avec_preferences_skyline_unifiees_min_min}, we consider, for convenience, two evaluation criteria having, in the first case, unified Skyline preferences $(\texttt{MIN}, \texttt{MIN})$, and in the second case, mixed Skyline preferences $(\texttt{MIN}, \texttt{MAX})$: with $\texttt{MIN}$ preference being X-axis, and $\texttt{MAX}$ preference in y-axis.
		
		Points $A$, $B$, $C$ and $D$ are not dominated. Whereas points $E$, $F$ and $G$ are not on the Pareto front (all colored segments) because they are dominated by other points. $A$, $B$, $C$ and $D$ are called efficient, and Pareto-optimal. The theoretical ideal point, or abstract, noted $I+$, with coordinates $(30, 30)$ on the left of figure~\ref{fig:cosinus_de_salton_avec_preferences_skyline_unifiees_min_min}, and with coordinates $(30, 80)$ on the right, dominates all Skyline points.
		
		The Salton cosine respectively for angles $\theta$ and $\theta'$, each formed by the vector from origin to $I+$ and the vector from origin to $D$, has an approximative value of $0.534~rad$ on the left of figure~\ref{fig:cosinus_de_salton_avec_preferences_skyline_unifiees_min_min}, and an approximative value of $0.611~rad$ on the right.
		
		Here, unification has been done using complement conversion (using \emph{supremum}) instead of inversion, for convenience. Indeed, when values are inverted (with some being lower than $1$ whereas other are much higher) representation becomes less readable. Nevertheless, in each case, the problematic remains.
	\end{example}
	
	\begin{example}
		To unify \texttt{Pokémon} relation Skyline preferences (cf. tableau~\ref{tab:relation_exemple}), as the $\texttt{MIN}$ preference is initially more frequent than the $\texttt{MAX}$ one, it is preferable to use $(\texttt{MIN}, \texttt{MIN}, \texttt{MIN})$ by inverting \texttt{Win} attribute's values (becoming \texttt{Win$^{-1}$} or \texttt{Loss}), as with table~\ref{tab:relation_exemple_4}.
	\end{example}
	
	\subsection{RankSky Method}\label{ssec:methode_RankSky}\index{RankSky}
	
	In this subsection, in order to propose a Skyline ranking solution to differentiate and order dissociated Skyline points, we present the adaptation of Google's well-known PageRank algorithm to ranking Skylines, RankSky (for PageRank Skyline). 
	
	More than the original version, poor in theoretical basis and presenting a value-based approach (\cite{pagePageRankCitationRanking1998}), we will focus on the rigorous linear algebraic framework to understand search engine ranking algorithms, with a particular work on Google’s PageRank mathematics, presenting a matrix-based approach (\cite{langvilleGooglesPageRankScience2006}). This method computes a column-stochastic matrix representing transition probabilities associated to Markov chains. The PageRank vector is defined as the principal eigenvector corresponding to the dominant eigenvalue $\lambda = 1$ of this matrix. Also, discussion about the role of the damping factor to ensure irreducibility and aperiodicity, guaranteeing a unique stationary distribution, is also of interest for our study.
	
	RankSky is a multi-step approach not using dominance relation or a time-consuming mathematical function as logarithm\footnote{Logarithm is a transcendental function way more costly, partly due to material optimization existing for adding and multiplying, than standard operations.}.
	
	It is regretfully not possible to embed this method in a relational DBMS.
	
	This computation is presented in details in this subsection, and the algorithm~\ref{algo:rank_sky} is proposed. For each step, we consider $i, j \in \{1, \dotsc, m\}$ (where $m$ is the Skyline tuple count).
	
	\begin{example}
		RankSky method can be applied on the \texttt{Pokémon} relation (cf. table~\ref{tab:relation_exemple_3}), with unified Skyline preferences $(\texttt{MAX}, \texttt{MAX}, \texttt{MAX})$. Indeed, for Google's PageRank ranking, it is mandatory to only use the preference $\texttt{MAX}$.
		
		\begin{table}[htbp]
			\caption{$\texttt{Pokémon}_{2}$ relation}\label{tab:relation_exemple_3}
			\centering
			\begin{minipage}{\linewidth}
				\centering
				\begin{tabular}{c|ccc}
					\toprule
					\texttt{RowId} & \texttt{Rarity}$^{-1}$ & \texttt{Duration}$^{-1}$ & \texttt{Win} \\
					\midrule
					$1$ & $1 / 5$ & $1 / 20$ & $70$ \\
					$2$ & $1 / 4$ & $1 / 60$ & $50$ \\
					$3$ & $1 / 5$ & $1 / 30$ & $60$ \\
					$4$ & $1 / 1$ & $1 / 80$ & $60$ \\
					$5$ & $1 / 5$ & $1 / 90$ & $40$ \\
					$6$ & $1 / 9$ & $1 / 30$ & $50$ \\
					$7$ & $1 / 7$ & $1 / 80$ & $40$ \\
					$8$ & $1 / 9$ & $1 / 90$ & $30$ \\
					\bottomrule
				\end{tabular}
			\end{minipage}
		\end{table}
	\end{example}
	
	\subsubsection{Step I: Making a square matrix}\label{sssec:etape_1_calcul_de_la_matrice_carree}
	
	In addition to a preference unification configuration step, the first two steps of CoSky method can be applied to let $S$ be, prior to that step, a normalized and ponderated matrix representation of a SkyLine where all criteria have a \texttt{MAX} preference.
	
	As RankSky Method is based on Google's well-known PageRank method, it requires a square matrix. We point out that a matrix multiplication of $M$ by its transpose $M^\intercal$ provides a square matrix $A$. Thus we have:
	
	\begin{equation}
		\forall M \in \mathbb{R}^{m \times n}, M^\intercal \in \mathbb{R}^{\intercal n \times m} \wedge M \cdot M^\intercal =
		\left\lbrace
		\begin{array}{l}
			A \in \mathbb{R}^{m \times m}, \forall m \geq n \\ 
			A \in \mathbb{R}^{n \times n}, \forall n \geq m
		\end{array}
		\right.
	\end{equation}
	
	We chose that solution to ensure a square matrix, moreover its computation cost is relatively low.
	
	Let $M$ be a matrix of a Skyline with unified Skyline preferences $(\texttt{MAX}, \dotsc, \texttt{MAX})$, and $\top$ the transpose of a matrix operator, thus we have $M^\intercal := \top(M)$ and :
	\begin{equation}
		A := M \cdot M^\intercal
	\end{equation}
	
	With $A$ a square matrix.
	
	\begin{example}
		\textcolor{new}{Here is} the matrix $M$ of the Skyline, with unified Skyline preferences $(\texttt{MAX}, \texttt{MAX}, \texttt{MAX})$, of the \texttt{Pokémon} relation.
		
		\begin{table}[htbp]
			\caption{Skyline's matrix $M$ of the $\texttt{Pokémon}_{2}$ relation}
			\centering
			\begin{minipage}{\linewidth}
				\centering
				\vspace{-1em}
				\begin{equation*}
					\setlength{\arraycolsep}{1em}
					M = 
					\begin{pmatrix}
						1/5 & 1/20 & 70 \\
						1/4 & 1/60 & 50 \\
						1/1 & 1/80 & 60 \\
					\end{pmatrix}
				\end{equation*}
			\end{minipage}
		\end{table}
		
		\textcolor{new}{Here is} the Skyline's transpose matrix $M^\intercal$ of the \texttt{Pokémon} relation.
		
		\begin{table}[htbp]
			\caption{Skyline's transpose matrix $M^\intercal$ of the $\texttt{Pokémon}_{2}$ relation}
			\centering
			\begin{minipage}{\linewidth}
				\centering
				\vspace{-1em}
				\begin{equation*}
					\setlength{\arraycolsep}{1em}
					M^\intercal = 
					\begin{pmatrix}
						1/5 & 1/4 & 1/1 \\
						1/20 & 1/60 & 1/80 \\
						70 & 50 & 60 \\
					\end{pmatrix}
				\end{equation*}
			\end{minipage}
		\end{table}
		
		\textcolor{new}{Here is} the square matrix $A = M \cdot M^\intercal$.
		
		\begin{table}[htbp]
			\caption{Square matrix $A = M \cdot M^\intercal$}
			\centering
			\begin{minipage}{\linewidth}
				\centering
				\vspace{-1em}
				\begin{equation*}
					\setlength{\arraycolsep}{1em}
					A = M \cdot M^\intercal \approx 
					\begin{pmatrix}
						4900 & 3500 & 4200 \\
						3500 & 2500 & 3000 \\
						4200 & 3000 & 3601 \\
					\end{pmatrix}
				\end{equation*}
			\end{minipage}
		\end{table}
	\end{example}
	
	\subsubsection{Step II: Making stochastic matrix}\label{sssec:etape_2_calcul_de_la_matrice_stochastique}
	
	A matrix $P \in \mathbb{R}^{m \times m}$ is called a row-stochastic matrix, left-stochastic matrix or simply stochastic matrix or Markov matrix\footnote{Such matrices are named Markov matrix because they are commonly used to represent transition probabilities in Markov chains.} if it satisfies the following conditions, with $\forall (i, j), p_{ij} \in P$, thus we have:
	\begin{itemize}
		\item $p_{ij} \geq 0$
		\item $\sum_{j=1}^{m} p_{ij} = 1$
	\end{itemize}
	
	Furthermore, as a consequence of Perron-Frobenius theorem, which applies to non-negative matrices, there always exists at least an eigenvector, or characteristic vector, $\mathbf{v} \in \mathbb{R}^n$ such that $P \cdot \mathbf{v} = \mathbf{v}$, \emph{i.e.}, 1 is always an eigenvalue, characteristic value, or characteristic root, $\lambda$ of $P$.
	
	Moreover, if $P$ is irreducible and aperiodic, then the eigenvalue $\lambda = 1$ is simple (\emph{i.e.} it has multiplicity one) and there exists a unique, up to scaling, positive eigenvector $\mathbf{v}$, called the stationary distribution, such that: $P \cdot \mathbf{v} = \mathbf{v}$ where $\sum_{i=1}^{n} v_i = 1$. 
	
	This property is required by Google's PageRank method. The transition matrix is modified by a teleportation, or damping factor, to guarantee these two properties, thus ensuring the existence and uniqueness of the eigenvector that represents PageRank scores.
	
	Let $A$ be a square matrix, $\forall (i, j), a_{ij} \in A$, thus we have $P$, a stochastic matrix, defined as follows:
	\begin{equation}
		p_{ij} = \frac{a_{ij}}{\sum_{i', j' = 1}^{m}a_{i'j'}}, \forall (i, j), p_{ij} \in P
	\end{equation} 
	
	\begin{example}
		\textcolor{new}{Here is} the stochastic matrix $P$ of $A$, having each line's sum equal to 1.
		
		\begin{table}[htbp]
			\caption{Stochastic matrix $P$ of $A$}
			\label{tab:matrice_p}
			\centering
			\begin{minipage}{\linewidth}
				\centering
				\vspace{-1em}
				\begin{equation*}
					\setlength{\arraycolsep}{1em}
					P \approx 
					\begin{pmatrix}
						4900/12600 & 3500/12600 & 4200/12600 \\
						3500/9000 & 2500/9000 & 3000/9000 \\
						4200/10801 & 3000/10801 & 3601/10801 \\
					\end{pmatrix}
				\end{equation*}
			\end{minipage}
		\end{table}
	\end{example}
	
	\subsubsection{Step III: Making Google's PageRank skyline matrix}\label{sssec:etape_3_calcul_de_la_matrice_pagerank_de_google}
	
	Let $A$ be a square and stochastic matrix representing a Skyline where all criteria have a \texttt{MAX} preference. We compute $A$ using the two previous steps. We can then calculate Google's PageRank skyline matrix $G$, column-stochastic, irreducible and aperiodic. In this context, $G$ is considered as a transition matrix associated to Markov chains, in which each line's sum is equal to 1.
	
	As a common practice (\cite{langvilleGooglesPageRankScience2006}), $\alpha = 0.85$ is a relevant damping factor so that, seeking the dominant eigenvector, the algorithm is efficient, \emph{i.e.}, it requires a limited iteration count.  
	
	Let $P$ be a stochastic matrix, $\alpha = 0.85$, a damping factor and $E$ a teleportation matrix such as $\forall (i, j), e_{ij} \in E \wedge e_{ij} = 1$, thus we have $G$, a Google PageRank skyline matrix, defined as follows:
	\begin{equation}
		G = \alpha \cdot P + \frac{1 - \alpha}{m} \cdot E
	\end{equation} 
	
	\begin{example}
		\textcolor{new}{Here is} the Google's PageRank skyline matrix $G$ with $\alpha = 0.85$.
		
		\begin{table}[htbp]
			\caption{Google's PageRank skyline matrix $G$ with $\alpha = 0.85$}
			\centering
			\begin{minipage}{\linewidth}
				\centering
				\vspace{-1em}
				\begin{equation*}
					\setlength{\arraycolsep}{1em}
					G \approx 0.85 \cdot  
					\begin{pmatrix}
						4900/12600 & 3500/12600 & 4200/12600 \\
						3500/9000 & 2500/9000 & 3000/9000 \\
						4200/10801 & 3000/10801 & 3601/10801 \\
					\end{pmatrix}
					+ 
					0.15/3 \cdot 
					\begin{pmatrix}
						1 & 1 & 1 \\
						1 & 1 & 1 \\
						1 & 1 & 1 \\
					\end{pmatrix}
				\end{equation*}
				\begin{equation*}
					\setlength{\arraycolsep}{1em}
					\approx
					\begin{pmatrix}
						0.38055071 & 0.28610903 & 0.33334026 \\
						0.38054699 & 0.2861075 & 0.33334551 \\
						0.38052694 & 0.28609908 & 0.33337398 \\
					\end{pmatrix}
				\end{equation*}
			\end{minipage}
		\end{table}
	\end{example}
	
	\subsubsection{Step IV: Score with IPL algorithm}\label{sssec:etape_4_score_avec_l_algorithme_ipl}
	
	Let $G$ be a Google's PageRank skyline matrix, $\epsilon = 10^{-p}$ the machine epsilon in term of $10^{-p}$, with precision $p$, $V_k = (1/m, \dotsc, 1/m)$ the row score eigenvector, we iteratively calculate $V_k = V_{k-1} \cdot G$ using IPL (Iterated Power-Like) algorithm until $|||V_k||_1 - ||V_{k-1}||_1| < \epsilon$, with $||V_x||_1$ the $l^1$ norm or Manhattan norm of vector $V_x$.
	
	IPL generalizes power iteration, power method, or von Mises iteration, improving convergence and adapting to particular matrix structures. The classical power iteration is a simple method to calculate the dominant eigenvector, corresponding to the largest eigenvalue in magnitude, therefore the maximal value, of a matrix. In RankSky, we modify recurrence formulas, that still preserve the direction of convergence, although not always monotone, toward the dominant eigenvector.
	
	IPL is relevant for matrices multiplications, including large but sparse matrices, like Google's PageRank calculation. The algorithm outputs a one-dimensional matrix, or 1-D matrix, which is the result of a matrix multiplication of a probability vector and a transition matrix.
	
	\textcolor{new}{We do not present the IPL algorithm implementation separately because it is integrated into Algorithm~\ref{algo:rank_sky}.}
	
	\begin{example}
		\textcolor{new}{Here is} the score vector $V_k$, which is the one-dimensional matrix, and eigenvector, provided by successive iterations of IPL algorithm on $G$.
		
		\begin{table}[htbp]
			\caption{Score vector $V_k$}
			\centering
			\begin{minipage}{\linewidth}
				\centering
				\vspace{-1em}
				\begin{equation*}
					\setlength{\arraycolsep}{1em}
					V_k \approx 
					\begin{pmatrix}
						0.3805417230797888 & 0.2861052759746371 & 0.33335300094557424 \\
					\end{pmatrix}
				\end{equation*}
			\end{minipage}
		\end{table}
	\end{example}
	
	\subsubsection{Step V: Sort outcomes}\label{sssec:etape_5_classement_des_resultats_de_ranksky}
	
	In the last step Skyline points are sorted by descending scores.
	
	\begin{example}
		The table~\ref{tab:classement_de_skyline_avec_la_methode_ranksky} portrays the \texttt{Pokémon} relation's Skyline ranked by RankSky method, with associated scores.
		
		\begin{table}[htb]
			\caption{Skyline ranking with RankSky}\label{tab:classement_de_skyline_avec_la_methode_ranksky}
			\centering
			\begin{minipage}{\linewidth}
				\centering
				\begin{tabular}{c|cccc}
					\toprule
					\texttt{RowId} & \texttt{Rarity} & \texttt{Duration} & \texttt{Win}$^{-1}$ & \texttt{Score} \\
					\midrule
					$1$ & $5$ & $20$ & $1 / 70$ & $0.381$ \\
					$4$ & $1$ & $80$ & $1 / 60$ & $0.333$ \\
					$2$ & $4$ & $60$ & $1 / 50$ & $0.286$ \\
					\bottomrule
				\end{tabular}
			\end{minipage}
		\end{table}
	\end{example}
	
	\subsubsection{RankSky Algorithm}
	
	The algorithm~\ref{algo:rank_sky} presents an RankSky Algorithm implementation, with skyline criteria preference unification, to \texttt{MAX}, as the only configuration step.
	
	\begin{algorithm}[htb]
		\caption{RankSky ($\mathcal{O}(|r| \log |r| + |S| \cdot |\mathcal{D}|^2 + k \cdot |S|^2)$)\label{algo:rank_sky}}
		\begin{algorithmic}
			\Require~ \\
			$r$ relation. \\
			Number of results $k$. \\
			\Ensure~ \\
			RankSky's ordered score table $score$. \\
			\State $S$: $r$'s Skyline
			\Comment{Computed with BBS}
			\State $\mathcal{D} := \{d_1, \dotsc, d_n \}$: $r$'s dimensions set.
			\State $M$: an array of $|S|$ arrays of $|\mathcal{D}|$ 0
			\State $M^\intercal$: an array of $|\mathcal{D}|$ arrays of $|S|$ 0
			\Comment{The transpose matrix}
			\State $A$: an array of $|S|$ arrays of $|S|$ 0
			\Comment{The square matrix}
			\State $P$: an array of $|S|$ arrays of $|S|$ 0
			\Comment{The stochastic matrix}
			\State $E$: an array of $|S|$ arrays of $|S|$ 1
			\Comment{The teleportation matrix}
			\State $G$: an array of $|S|$ arrays of $|S|$ 0
			\Comment{The Google matrix}
			\State $\alpha := 0.85$;
			\Comment{The $\alpha$ factor}
			\State $p := 3$;
			\Comment{The calculation precision}
			\State $V_k$: an array of $|S|$ $1 / |S|$
			\State $Z_k$: an array of $|S|$ 0
			\State $||V_k||_1$: a positive integer
			\State $||Z_k||_1 := 0$:
			\State $score$: an associative array of $|\mathcal{D}| + 1$ 0
			\State
			\Comment{Step I: Making $M$ a square matrix}
			\ForAll{$t \in S$}
			\State $i := t[\texttt{RowId}]$;
			\ForAll{$d_j \in \mathcal{D}$}
			\State $M[i][d_j] := t[d_j]$;
			\State $score[i][d_j] := t[d_j]$;
			\EndFor
			\EndFor
			\State $M^\intercal := \top(M)$;
			\Comment{With $\top(M)$: $M \in \mathbb{R}^{m \times n} \to M^\intercal \in \mathbb{R}^{n \times m}$}
			\State $A := M \cdot M^\intercal$;
			\State
			\Comment{Step II: Making $A$ a stochastic matrix}
			\State $P = A$;
			\For{$i := 0, \dots, |P| - 1$}
			\State $tot := 0$;
			\For{$j := 0, \dots, |P| - 1$}
			\State $tot := tot + P[i][j]$;
			\EndFor
			\For{$j := 0, \dots, |P| - 1$}
			\State $P[i][j] := P[i][j] / tot$;
			\EndFor
			\EndFor
			\State
			\Comment{Step III: Making Google's PageRank skyline matrix}
			\State $G := \alpha \cdot P + (1 - \alpha) / |P| \cdot E$;
			\State
			\Comment{Step IV: Score with IPL (Iterated Power-Like) algorithm}
			\State $\epsilon := 10^{-p}$;
			\Comment{The machine epsilon in term of $10^{-p}$, with precision $p$}
			\State $||Z_k||_1 := 0$;
			\Loop
			\State $||V_k||_1 := ||Z_k||_1$;
			\State $Z_k := V_k \cdot G$;
			\State $||Z_k||_1 := 0$;
			\For{$i := 0, \dots, |Z_k| - 1$}
			\State $||Z_k||_1 := ||Z_k||_1 + |Z_k[i]|$;
			\Comment{$l^1$ norm or Manhattan norm: $||Z_k||_1 = \sum_{i=1}^{n} |Z_{k_i}|$}
			\EndFor
			\State $V_k := Z_k / ||Z_k||_1$;
			\If{$|||Z_k||_1 - ||V_k||_1| < \epsilon$}
			\State \textbf{break}
			\EndIf
			\EndLoop
			\State
			\Comment{Step V: Sort outcomes}
			\ForAll{$t \in V_k$}
			\State $i := t[\texttt{RowId}]$;
			\State $score[i][|\mathcal{D}|] := V_k[i]$;
			\EndFor
			\State $score := reverseQuickSort(score, 0, |S| - 1)$;
			\State
			\Comment{$score$ descending sort on $score[i][|\mathcal{D}|]$ values, with $\forall t \in S, i := t[\texttt{RowId}]$}
			\Return $score$
		\end{algorithmic}
	\end{algorithm}

	\color{new}

	\subsubsection{Strengths and limitations}
	
	By its very nature, the RankSky method pretty shares the same advantages and inconveniences as the well-known PageRank method.
	
	Thus, RankSky captures global importance. It considers not only how many associations a point has, but also the importance of those linking to it. RankSky propagates influence, meaning a point is important if important points have a path to it. Unlike naive ranking, this method is resistant to simple manipulation because it is harder to exploit. It considers which points are associated, not just how many. RankSky is scalable for both sparse and large datasets and uses efficient iterative computation. Skyline points without outgoing edges, or dangling points, are still handled in a principled way using teleportation.
	
	However, for the same reason, the RankSky method is also insensitive to content or context as it only considers the data structure and not the content of the Skyline points or user intent. Therefore, a high-ranked point may be irrelevant to a specific query or topic. This method is also a slow-to-update method that is computationally intensive, especially for very large datasets. It is not ideal for frequently changing needs unless they are optimized or approximated. In a shifting environment, the method presents biases toward older or regarded as important Skyline points. These points dominate the ranking, which can lead to bias against new points or the overlooking of isolated but high-quality points.
	
	Sure enough, this solution requires careful parameter tuning. Indeed, the damping factor, which is usually set to $0.85$, affects the convergence and fairness of the ranking. Thus, poor tuning can impact outcomes or convergence speed.
	
	The RankSky method could be improved by integrating a personalized PageRank variant to propose customization that provides no longer just a global ranking, even if this approach is more complex and expensive and must therefore be reserved for specific uses.
	
	\color{black}
	
	\subsection{CoSky Method}\label{ssec:methode_CoSky}\index{CoSky}
	
	In order to propose an alternative Skyline ranking solution to differentiate and order dissociated Skyline points, we now present the CoSky method. CoSky (for cosine Skyline) is also a multi-step approach not using dominance relation, logarithm\footnote{Most programming mathematical libraries implement, for the cosine function, a angle reduction follows by a fast series expansion like Taylor's, and for the logarithmic function, a more involved process with more steps and conditionals like an extraction of the exponent and mantissa from the floating-point representation follows by a computation of the logarithm of the mantissa via interpolation or polynomial approximation, once more completed by a result fusion.} and matrix calculation. To our knowledge, it is the first TOPSIS method\footnote{TOPSIS for Technic for Order Preference by Similarity to Ideal Solution, is a method for ranking alternatives on a given order based on beneficial or unfavorable criteria.} (\cite{laiTOPSISMODM1994}, \cite{behzadianStateofTheartSurvey2012}) applied to this type of ranking. TOPSIS is based on a vectorial normalization, a weighting computation for each attribute, and a score computation for each point determined by a geometrical measure of distances between each alternative, represented by a point, and the ideal/anti-ideal solutions. In CoSky, attributes normalization is done with sum, an automatic normalized attributes weighting with Gini index, and the score with Salton cosine of the angle between a Skyline point and the ideal point.
	
	This computation is presented in details in this subsection. For each step, we consider $i \in \{1, \dotsc, m\}$ and $j \in \{1, \dotsc, n\}$ (where $m$ is the tuple count and $n$ the attribute count).
	
	\begin{example}
		CoSky method can be applied on the \texttt{Pokémon} relation (cf. table~\ref{tab:relation_exemple_4}), with unified Skyline preferences $(\texttt{MIN}, \texttt{MIN}, \texttt{MIN})$.
		
		\begin{table}[htbp]
			\caption{$\texttt{Pokémon}_{3}$ relation}\label{tab:relation_exemple_4}
			\centering
			\begin{minipage}{\linewidth}
				\centering
				\begin{tabular}{c|ccc}
					\toprule
					\texttt{RowId} & \texttt{Rarity} & \texttt{Duration} & $\texttt{Win}^{-1}$ \\
					\midrule
					$1$ & $5$ & $20$ & $1 / 70$ \\
					$2$ & $4$ & $60$ & $1 / 50$ \\
					$3$ & $5$ & $30$ & $1 / 60$ \\
					$4$ & $1$ & $80$ & $1 / 60$ \\
					$5$ & $5$ & $90$ & $1 / 40$ \\
					$6$ & $9$ & $30$ & $1 / 50$ \\
					$7$ & $7$ & $80$ & $1 / 40$ \\
					$8$ & $9$ & $90$ & $1 / 30$ \\
					\bottomrule
				\end{tabular}
			\end{minipage}
		\end{table}
	\end{example}
	
	\subsubsection{Step I: Normalization by sum}\label{sssec:etape_1_normalisation_des_attributs_par_la_somme}
	
	Skyline is normalized by using sum. This method assures all normalized values are between $-1$ and $1$, with a sum $1$ by dividing each value by the sum of all values. It is well suited when data relative scale is more important than their absolute values, and when proportion to the total are relevant, can remove anomalies, notably coming from different measure units or scales, while ensuring measurability and comparability of attributes. Let $x$ be a value and $\Sigma x$ the sum of all values, then $x$'s normalized value $x'$ is:
	\[x' = \frac{x}{\Sigma x}\]
	
	It is advised to prior transpose all values to natural real numbers. Moreover, this way every attribute value of every Skyline point is converted into a $0$ to $1$ value.
	
	Let $S_N$ be a Skyline normalized points set, or normalized Skyline, and $u_i = (u_{i}[A_1], u_{i}[A_2], \dotsc, u_{i}[A_n]) \in S_N$ a tuple, thus we have:
	\begin{equation}
		u_{i}[A_j] = \frac{t_i [A_j]}{\sum_{i' = 1}^{m}t_{i'}[A_j]}, \forall t_i \in S
	\end{equation} 
	
	Beware, this method implies a divide by $0$ risk, on attributes non normalized values' sum: $\sum_{i' = 1}^{m}t_{i'}[A_j] \neq 0$.
	
	\begin{example}
		The table~\ref{tab:skyline_normalise_par_la_somme} portrays the \texttt{Pokémon} relation's Skyline normalized by sum.
		
		\begin{table}[htbp]
			\caption{\texttt{Pokémon} relation's Skyline normalized by using sum}\label{tab:skyline_normalise_par_la_somme}
			\centering
			\begin{minipage}{\linewidth}
				\centering
				\begin{tabular}{c|ccc}
					\toprule
					\texttt{RowId} & \texttt{Rarity} & \texttt{Duration} & $\texttt{Win}^{-1}$ \\
					\midrule
					$1$ & $5/10$ & $20/160$ & $1 / 70 \times 2100 / 107$ \\
					$2$ & $4/10$ & $60/160$ & $1 / 50 \times 2100 / 107$ \\
					$4$ & $1/10$ & $80/160$ & $1 / 60 \times 2100 / 107$ \\
					\bottomrule
				\end{tabular}
				
				Or, with approximate values:
				
				\begin{tabular}{c|ccc}
					\toprule
					\texttt{RowId} & \texttt{Rarity} & \texttt{Duration} & $\texttt{Win}^{-1}$ \\
					\midrule
					$1$ & $0.5$ & $0.125$ & $0.280$ \\
					$2$ & $0.4$ & $0.125$ & $0.393$ \\
					$4$ & $0.1$ & $0.5$ & $0.327$ \\
					\bottomrule
				\end{tabular}
			\end{minipage}
		\end{table}
	\end{example}
	
	\subsubsection{Step II: Gini index-based weighting}\label{sssec:etape_2_ponderation_automatique_des_attributs_normalises_avec_l_indice_de_gini}
	
	To strictly distinguish Skyline points, it is essential to set up a discriminating measure, and several have been proposed in the literature. An entropy based method was used for multi criteria problems (\cite{huangCombiningEntropyWeight2008}, \cite{lotfiImpreciseShannonsEntropy2010}). This method is well suited for Skylines, where decision making requires a simple comparison of attributes values. However, as entropy computation needs a time-consuming logarithm function, this method is inefficient.
	
	We would rather set up another measure, the Gini index\footnote{The Gini index (or coefficient) is a statistical measure used to evaluate outliers variables in a given population. It is mainly used to measure a country's salary inequalities. It spans between $0$ (full correlation) and $1$ (no correlation), where the higher the index, the greater the inequality}, while being without logarithm, it is faster to compute than the entropy method. In the presented method, the Gini index is used to derive the attributes' weight in order to establish a divergence degree of attributes' values. Gini index $A_j$, $Gini(A_j)$, is calculated by the following equation:
	\begin{equation}
		Gini(A_j) = 1 - \sum_{i = 1}^{m}u_{i}[A_j]^2
	\end{equation}
	
	We call $W$ the attribute $A_j$'s weight. $(W(A_1), W(A_2), \dotsc, W(A_m))$ can be specified by the decision-maker so that $\texttt{SUM}(W(A_1), W(A_2), \dotsc, W(A_m)) = 1$. The attribute's weight is defined as follows:
	\begin{equation}
		W(A_j) = \frac{Gini(A_j)}{\sum_{j' = 1}^{n} Gini(A_{j'})} 
	\end{equation}
	
	Let $S_P$ be a weighted Skyline, or Skyline weighted points set, and $v_i  = (v_{i}[A_1], v_{i}[A_2], \dotsc, v_{i}[A_n]) \in S_P$ a tuple, thus we have:
	\begin{equation}
		v_{i}[A_j] = W (A_j) \times u_{i}[A_j], \forall u_i \in S_N
	\end{equation}
	
	\begin{example}
		The table~\ref{tab:skyline_normalise_pondere_avec_indice_de_gini} portrays the \texttt{Pokémon} relation's Skyline normalized weighted by Gini's index.
		
		\begin{table}[htbp]
			\caption{\texttt{Pokémon} relation's Skyline normalized weighted by Gini's index}\label{tab:skyline_normalise_pondere_avec_indice_de_gini}
			\centering
			\begin{minipage}{\linewidth}
				\centering
				\begin{tabular}{c|ccc}
					\toprule
					\texttt{RowId} & \texttt{Rarity} & \texttt{Duration} & $\texttt{Win}^{-1}$ \\
					\midrule
					$1$ & $0.158$ & $0.041$ & $0.101$ \\
					$2$ & $0.126$ & $0.121$ & $0.141$ \\
					$4$ & $0.032$ & $0.162$ & $0.118$ \\
					\bottomrule
				\end{tabular}
			\end{minipage}
		\end{table}
	\end{example}
	
	\subsubsection{Step III: Determination of the ideal point}\label{sssec:etape_3_determination_du_point_ideal}
	
	The theoretical, or abstract, ideal point, noted $I^+$, dominating every Skyline points, is a tuple that optimally matches all Skyline preferences.
	
	Thus, with $I^+ = (I^+[A_1], I^+[A_2], \dotsc, I^+[A_n])$, we have:
	\begin{equation}
		I^+[A_j] = 
		\left\lbrace
		\begin{array}{l}
			\texttt{MAX}(v_i[A_j]) \equiv Pref(A_j) = \texttt{MAX} \\ 
			\texttt{MIN}(v_i[A_j]) \equiv Pref(A_j) = \texttt{MIN} 
		\end{array}
		\right.
	\end{equation}
	
	\begin{example}
		With the \texttt{Pokémon} relation (cf. table~\ref{tab:relation_exemple_4}), the ideal Pokemon sequence in a fight combines conditions on the lowest possible \texttt{Rarity}, on the shortest possible \texttt{Duration}, and on the lowest possible \texttt{Win}$^{-1}$ rate. The ideal attributes approximate values compute by CoSky method are:
		\begin{itemize}
			\item \texttt{Rarity}: $0.032$
			\item \texttt{Duration}: $0.040$
			\item \texttt{Win}$^{-1}$: $0.101$
		\end{itemize}
	\end{example}
	
	\subsubsection{Step IV: Scores with Salton cosine}\label{sssec:etape_4_calcul_des_scores_avec_le_cosinus_de_salton}
	
	This step computes a Skyline point's score using Salton cosine \footnote{Salton cosine, or similarity measure cosine, or similarity cosine, measures, between 0 à 1, the similarity between vectors. A piece of information can be represented by a vector and its value by an angle in a vector space. It is traditionally used by search engines to rank Web pages.}. To that end, we compute the cosine of the angle between the ideal point and the Skyline point. The narrower the angle (therefore, the higher the angle's cosine), the more important the Skyline point is.
	
	Let $S_{Score}$ be a set of all Skyline point score, $v_i  = (v_{i}[A_1], v_{i}[A_2], \dotsc, v_{i}[A_n]) \in S_P$ a tuple and $I^+ = (I^+[A_1], I^+[A_2], \dotsc, I^+[A_n])$ the ideal point, thus we have:
	\begin{equation}
		s_i = S_c(v_i, I^+) := cos(\theta) = \frac{v_i \cdot I^+}{||v_i|| \cdot ||I^+||}
	\end{equation}
	\begin{equation}\label{eq:cosinus_de_salton}
		s_i = \frac{\sum_{j=1}^n v_{i}[A_j] \cdot I^+[A_j]}{\sqrt{\sum_{j=1}^n v_{i}[A_j]^2} \cdot \sqrt{\sum_{j=1}^n I^+[A_j]^2}}, \forall s_i \in S_{Score}
	\end{equation}
	
	In consequence, $s_i = 1$ if and only if the Skyline point is considered as the most interesting one, and respectively $s_i = 0$ if and only if it is considered as the least interesting.
	
	We can use TOPSIS' similarity principle to compute each Skyline point's score as follows: let $I^-$ be the anti-ideal point then, $\forall v_i \in S_P$, if we consider $I^- = (I^-[A_1], I^-[A_2], \dotsc, I^-[A_n])$, we have:
	\begin{equation}
		I^-[A_j] = 
		\left\lbrace
		\begin{array}{l}
			\texttt{MAX}(v_i[A_j]) \equiv Pref(A_j) = \texttt{MIN} \\ 
			\texttt{MIN}(v_i[A_j]) \equiv Pref(A_j) = \texttt{MAX}
		\end{array}
		\right.
	\end{equation}
	
	\subsubsection{Step V: Sort outcomes}\label{sssec:etape_5_classement_des_resultats_de_cosky}
	
	In the last step, as for RankSky, Skyline points are sorted by descending scores.
	
	\begin{example}
		The table~\ref{tab:classement_de_skyline_avec_la_methode_cosky} portrays the \texttt{Pokémon} relation's Skyline ranked by CoSky method, with associated scores.
		
		CoSky's Salton cosine gives different scores to each Skyline point, like RankSky but unlike dp-idp. Skyline points with \texttt{RowId} $2$ and $4$ have, with CoSky, differents non null scores, despite having a $0$ score with dp-idp. In this example, not only are Skyline point with \texttt{RowId} $2$ and $4$ differentiated, but also the sorting is different than RankSky and dp-idp.
		
		\begin{table}[htb]
			\caption{Skyline ranking with CoSky}\label{tab:classement_de_skyline_avec_la_methode_cosky}
			\centering
			\begin{minipage}{\linewidth}
				\centering
				\begin{tabular}{c|cccc}
					\toprule
					\texttt{RowId} & \texttt{Rarity} & \texttt{Duration} & \texttt{Win}$^{-1}$ & \texttt{Score} \\
					\midrule
					$2$ & $4$ & $60$ & $1 / 50$ & $0.909$ \\
					$4$ & $1$ & $80$ & $1 / 60$ & $0.847$ \\
					$1$ & $5$ & $20$ & $1 / 70$ & $0.774$ \\
					\bottomrule
				\end{tabular}
			\end{minipage}
		\end{table}
	\end{example}
	
	\subsubsection{CoSky in SQL}
	
	CoSky method is fully integrable into relational database management systems (DBMS), and can be expressed in a SQL query.
	
	\begin{example}
		CoSky method can be applied using the following SQL query\footnote{If the Skyline is composed by only one point, its ranking is useless or shall be done using $\texttt{COALESCE}(\texttt{NULLIF}(..., 0), 1)$ on each query denominator.}:
		\lstset{style=SQLStyle}
		\begin{lstlisting}[ language=SQL,
							deletekeywords={IDENTITY},
							deletekeywords={[2]INT},
							morekeywords={CLUSTERED, SKYLINE, OF, SQRT, ROUND, WITH},
							framesep=8pt,
							xleftmargin=40pt,
							framexleftmargin=40pt,
							frame=tb,
							framerule=0pt ]
			WITH S AS (
				SELECT *
				FROM Pokémon
				SKYLINE OF Rarity MIN, Duration MIN, Loss MIN
			), SN AS (
				SELECT RowId,
					   Rarity / TRarity AS NRarity,
					   Duration / TDuration AS NDuration,
				       Loss / TLoss AS NLoss 
				FROM S, 
				(
					SELECT SUM(Rarity) AS TRarity, 
						   SUM(Duration) AS TDuration, 
						   SUM(Loss) AS TLoss FROM S
				) AS ST
			), SGini AS (
				SELECT 1 - SUM(NRarity * NRarity) AS GRarity,
				       1 - SUM(NDuration * NDuration) AS GDuration,
				       1 - SUM(NLoss * NLoss) AS GLoss
				FROM SN
			), SW AS (
				SELECT GRarity / (GRarity + GDuration + GLoss) AS WRare,
				       GDuration / (GRarity + GDuration + GLoss) AS WDuration,
				       GLoss / (GRarity + GDuration + GLoss) AS WLoss
				FROM SGini
			), SP AS (
				SELECT RowId,  
					   WRare * NRarity AS PRarity,
				       WDuration * NDuration AS PDuration,
				       WLoss * NLoss AS PLoss
				FROM SN, SW
			), Idéal AS (
				SELECT MIN(PRarity) AS IRarity,  
				       MIN(PDuration) AS IDuration, 
				       MAX(PLoss) AS ILoss
				FROM SP
			), SScore AS (
				SELECT RowId,
				       (IRarity * PRarity + IDuration * PDuration + ILoss * PLoss) / 
				 	   (
					       SQRT(PRarity * PRarity + 
						        PDuration * PDuration + 
					            PLoss * PLoss) *
						   SQRT(IRarity * IRarity + 
							    IDuration * IDuration + 
					            ILoss * ILoss)
					   )
				AS Score 
				FROM Idéal, SP
			)
			SELECT P.RowId AS RowId, Rarity, Duration, Loss, ROUND(Score, 3) AS Score
			FROM S P 
			INNER JOIN SScore rs ON P.RowId = rs.RowId
			ORDER BY Score DESC;
		\end{lstlisting}
		
		And, the associated query without the Skyline operator is scematically depicted as follows:
		\lstset{style=SQLStyle}
		\begin{lstlisting}[ language=SQL,
							deletekeywords={IDENTITY},
							deletekeywords={[2]INT},
							morekeywords={CLUSTERED, SKYLINE, OF, SQRT, ROUND, WITH},
							framesep=8pt,
							xleftmargin=40pt,
							framexleftmargin=40pt,
							frame=tb,
							framerule=0pt ]
			WITH S AS (
			    SELECT *
		 	    FROM Pokémon AS P1
				WHERE NOT EXISTS (
					SELECT *
					FROM Pokémon AS P2
					WHERE (P2.Rarity <= P1.Rarity
					  AND P2.Duration <= P1.Duration
					  AND P2.Win >= P1.Win)
					  AND (P2.Rarity < P1.Rarity
					   OR P2.Duration < P1.Duration
					   OR P2.Win > P1.Win))
			), SN AS (
			    ...
			)
			...
			SELECT P.RowId AS RowId, Rarity, Duration, Loss, ROUND(Score, 3) AS Score
			FROM S P 
			INNER JOIN SScore rs ON P.RowId = rs.RowId
			ORDER BY Score DESC;
		\end{lstlisting}
	\end{example}
	
	\subsubsection{CoSky Algorithm}
	
	Sometimes, a code integrable algorithm can be more useful than a relational DBMS integrated one, as it is the case with a high dimensional cardinality, or as part of a library.
	
	To this end, we propose the naive algorithm\footnote{CoSky algorithm has been optimized, but without use of data statistics, parallelism, cache management, etc.}~\ref{algo:CoSky_1_2}.
	
	CoSky algorithm does not really compute the Skyline. It is, therefore, needed to use an external solution. Thus, we chose to used branch-and-bound Skyline (BBS) (\cite{papadiasProgressiveSkylineComputation2005}) for its efficiency.
	
	CoSky algorithm is, obviously, close to the CoSky SQL implementation. Method steps are conserved even if, for optimization purpose, their instructions and order can be altered.
	
	For readability purpose, algorithm has been documented using the \texttt{Pokémon} relation example (cf. table~\ref{tab:relation_exemple_4}), with unified Skyline preferences $(\texttt{MIN}, \texttt{MIN}, \texttt{MIN})$.
	
	\begin{algorithm}[htbp]
		\caption{CoSky ($\mathcal{O}(|S| \cdot |\mathcal{D}| + |r| \cdot \log(|r|))$)\label{algo:CoSky_1_2}}
		\begin{algorithmic}
			\Require~ \\
			$r$ relation. \\
			\Ensure~ \\
			CoSky's ordered score table $score$. \\
			\State $S$: $r$'s Skyline
			\Comment{Computed with BBS}
			\State $\mathcal{D} := \{d_1, \dotsc, d_n \}$: $r$'s dimensions set.
			\State $sum\mathcal{D}_S$: an associative array of $|\mathcal{D}|$ 0
			\State $S_N$: the future Skyline $S$ normalized
			\State $sum\mathcal{D}_{S_N}^2$: an associative array of $|\mathcal{D}|$ 0
			\State $gini$: a new associative array of $|\mathcal{D}|$ integers
			\State $sum_{gini} := 0$;
			\State $S_{NP}$: the future Skyline $S$ normalized and weighted
			\State $sum\mathcal{D}_{S_{NP}}^2$: an associative array of $|\mathcal{D}|$ 0
			\State $ideal$: an associative array of $|\mathcal{D}|$ 1
			\State $sum_{ideal}^2 := 0$;
			\State $sqrt_{sum_{ideal}^2}$: a positive integer
			\State $score$: an associative array of $|\mathcal{D}| + 1$ 0
			\ForAll{$t \in S$}
			\State $i := t[\texttt{RowId}]$;
			\ForAll{$d_j \in \mathcal{D}$}
			\State $sum\mathcal{D}_S[d_j] := sum\mathcal{D}_S[d_j] + t[d_j]$;
			\Comment{\emph{i.e.} SUM(Duration)}
			\State $score[i][d_j] := t[d_j]$;
			\EndFor
			\EndFor
			\ForAll{$t \in S_N$}
			\State $u$: $u \in S, u[\texttt{RowId}] = t[\texttt{RowId}]$;
			\ForAll{$d_j \in \mathcal{D}$}
			\State $t[d_j] := u[d_j] / sum\mathcal{D}_S[d_j]$;
			\Comment{\emph{i.e.} Duration / TDuration}
			\State $sum\mathcal{D}_{S_N}^2[d_j] := sum\mathcal{D}_{S_N}^2[d_j] + t[d_j]^2$;
			\Comment{\emph{i.e.} SUM(NDuration * NDuration)}
			\EndFor
			\EndFor
			\ForAll{$d_j \in \mathcal{D}$}
			\State $gini[d_j] := 1 - sum\mathcal{D}_{S_N}^2[d_j]$;
			\Comment{\emph{i.e.} 1 - SUM(NDuration * NDuration)}
			\State $sum_{gini} := sum_{gini} + gini[d_j]$;
			\Comment{\emph{i.e.} GRarity + GDuration + GLoss}
			\EndFor
			\ForAll{$t \in S_{NP}$}
			\State $i := t[\texttt{RowId}]$;
			\State $u$: $u \in S_N, u[\texttt{RowId}] = i$;
			\ForAll{$d_j \in \mathcal{D}$}
			\State $t[d_j] := gini[d_j] / sum_{gini} \times u[d_j]$;
			\State 
			\Comment{\emph{i.e.} GDuration / (GRarity + GDuration + GLoss)\dots}
			\State 
			\Comment{\dots \  AS WRarity and WDuration * NDuration}
			\State $sum\mathcal{D}_{S_{NP}}^2[i] := sum\mathcal{D}_{S_{NP}}^2[i] + t[d_j]^2$;
			\State 
			\Comment{\emph{i.e.} PRarity * PRarity + \dots \ + PLoss * PLoss}
			\If{$t[d_j] < ideal[d_j]$}
			\State $ideal[d_j] := t[d_j]$;
			\Comment{\emph{i.e.} MIN(Duration)}
			\EndIf
			\EndFor
			\EndFor
			\ForAll{$d_j \in \mathcal{D}$}
			\State $sum_{ideal}^2 := sum_{ideal}^2 + ideal[d_j]^2$; 
			\Comment{\emph{i.e.} IRarity * IRarity + \dots \ + ILoss * ILoss}
			\EndFor
			\State $sqrt_{sum_{ideal}^2} := \sqrt{sum_{ideal}^2}$;
			\ForAll{$t \in S_{NP}$}
			\State $i := t[\texttt{RowId}]$;
			\State $score_{numerator} := 0$;
			\ForAll{$d_j \in \mathcal{D}$}
			\State $score_{numerateur} := score_{numerateur} + ideal[d_j] \times t[d_j]$;
			\State 
			\Comment{\emph{i.e.} IRarity * PRarity + \dots \ + ILoss * PLoss}
			\EndFor
			\State $score[i][|\mathcal{D}|] := score_{numerateur} / (\sqrt{sum\mathcal{D}_{S_{NP}}^2[i]} \times sqrt_{sum_{ideal}^2})$;
			\Comment{cf. formula~\ref{eq:cosinus_de_salton}}
			\EndFor
			\State $score := reverseQuickSort(score, 0, |S| - 1)$;
			\State
			\Comment{$score$ descending sort on $score[i][|\mathcal{D}|]$ values, with $\forall t \in S, i := t[\texttt{RowId}]$}
			\Return $score$;
		\end{algorithmic}
	\end{algorithm}
		
	\color{new}
	
	\subsection{Strengths and limitations}
	
	By its very nature, the CoSky method shares many of the same advantages and inconveniences as the The Salton cosine, backed by a rich body of literature in IR and form the basis for many benchmarks and retrieval models.
	
	Thus, CoSKy measures the angle between vectors, not their magnitude, which allows comparison of Skyline points that may have dimension values with large deviations, but that are comparable in importance. This solution only requires dot products and vector norms, both of which are efficiently computable and scalable to large datasets. For non-negative vectors, the similarity score lies in the range of $[0, 1]$, with $1$ indicating identical orientation. These scores are easy to interpret, which makes them useful for analysis and evaluation. CoSky is domain-agnostic and can be applicable wherever data can be represented as vectors.
	
	However, for the same reason, the CoSky method relies on raw vector alignment and does not capture semantic similarity. It assumes a linear vector space structure. In highly sparse vector spaces, the Salton cosine may be undefined or uninformative, especially in the case of Skyline composed by few points. Therefore, in that case, CoSky does not really provide exploitable information. As CoSky lacks a statistical foundation, which makes it less interpretable in probabilistic terms.
	
	\color{black}
	
	\subsection{Top-$k$ method}\index{Top-$k$}
	
	For this last method, we use the multilevel Skyline principle (\cite{preisingerLookingBestNot2015}) to find the top-$k$ Skyline points, unordered. A top-$k$ $Q_k$ Skyline query on a relation $r$ computes top-$k$ points, based on $S$' Skyline preferences. Let $S_0(r)$ be the $0$ level points from the multilevel Skyline, or Skyline points, such as $S_0(r) = S$, and $Card(r)$ $r$'s cardinality such as $Card(r) > k$, then:
	\begin{itemize}
		\item if $Card(S_0(r)) > k$: $Q_k$ only returns $k$ points from $S_0(r)$;
		\item if $Card(S_0(r)) = k$: $Q_k$ returns the whole Skyline (\emph{i.e.} all points from $S_0(r)$);
		\item if $Card(S_0(r)) < k$: $S_0(r)$ do not have enough points to compute a valid answer with $Q_k$. A multilevel Skyline approach shall be applied. Thus, not only $S_1(r)$ points from $(r \backslash S_0(r))$ are returned, but also possibly $S_2(r)$ points from $(r \backslash (S_0(r) \cup S_1(r))$, from $S_3(r)$\dots as long as the cumulative result count is lower than $k$. 
	\end{itemize}
	
	\subsubsection{DeepSky}\index{DeepSky}
	
	The algorithm~\ref{algo:deep_sky}, DeepSky, uses this multilevel principle allied to a skyline ranking solution such as CoSky to find top-$k$ ranked Skyline points. It returns $k$ multilevel Skyline points with $k$ highest scores computed by such a skyline ranking solution.
	
	\begin{algorithm}[htb]
		\caption{DeepSky ($\mathcal{O}(k \cdot (|S| \cdot |\mathcal{D}| + |r| \cdot \log(|r|)))$)\label{algo:deep_sky}}
		\begin{algorithmic}
			\Require~ \\
			Algorithm for ranking Skyline $\texttt{<Algo>}$. \\
			$r$ relation. \\
			Number of results $k$. \\
			\Ensure~ \\
			Top-$k$ tuples/points with $top_k$ highest scores. \\
			\State $top_k := \emptyset$;
			\State $tot := 0$;
			\Comment{Total computed results count}
			\State $r_l := r$;
			\Comment{Current level}
			\While{$tot < k \vee r_l = \emptyset$}
			\State $S := \texttt{<Algo>}(r_l)$;
			\Comment{$\texttt{<Algo>}$ could be $\texttt{CoSky}$, $\texttt{RankSky}$ or $\texttt{dp-idp}$}
			\State $tot := tot + |S|$;
			\If{$tot \le k$}
			\State $top_k := top_k \cup S$;
			\State $r_l := r_l \backslash S$;
			\Else
			\State $S_trunc$: $S$' $k$ firsts points
			\State $top_k := top_k \cup S_trunc$;
			\Return $top_k$
			\EndIf
			\EndWhile
			\Return $top_k$
		\end{algorithmic}
	\end{algorithm}
	
	\begin{example}
		With \texttt{Pokémon} relation (cf. table~\ref{tab:relation_exemple_4}), and $k = 4$, DeepSky algorithm (cf. algorithm~\ref{algo:deep_sky}) returns Skyline points with \texttt{RowId} $1$, $4$ and $2$, those at level $0$, and with \texttt{RowId} $3$, the only level $1$ point.
	\end{example}
	
	\section{Discussion}\label{sec:discussion}
	
	\textcolor{new}{RankSky is an original method of ranking Skylines, inspired by the widely studied PageRank method. However, it may be more efficient in some cases to adapt the use of IPL to improve the convergence rate of the method when considering the dominant eigenvalue and the spectral ratio.}
	
	\textcolor{new}{CoSky is an original method of ranking skylines that uses the flexible, easy-to-compute, and easy-to-interpret Salton cosine method.} It is also invariant to vectors' linear transformations (such as values scaling) and is well fitted to compare vectors in high dimension spaces. However, it is weak against null vectors and does not take into account vectors' magnitude, only their direction. Thus, in the rare occasion where a Skyline point vector is overlapping the ideal point vector, with a distinct magnitude, it will be considered, wrongly, as optimal. Likewise, several Skyline points can have the same similarity measure.
	
	We assume an ideal point as optimal on the whole dimension set. We could consider, instead, a closer theoretical point dominating all Skyline points.
	
	\textcolor{new}{The method chosen depends on the context and the meaning the user wants to convey through the ranking.} With dp-idp, a point's importance is inversely proportional to the count of Skyline points dominating it, which can be detrimental in some cases, like with a dense cluster of points close to the Pareto front. RanKSky, uses a well-tested ranking method, yet sensitive to the needed Skyline preferences unification, which can contribute to roundoff errors, thus possibly altering ranking Skyline in case of high cardinality. CoSky, in turn, measures distance with an ideal, that always makes sense. 
	
	Other Skyline points ranking methods have been proposed in the literature with the same aim to reduce result sets. Among these, we could compare with, even if only experimentally, the method using regret minimisation (\cite{fabrisFlexibleSkylinesRegret2022}).
	
	\section{Experimental evaluations}\label{sec:evaluations_experimentales}
	
	The experimental evaluation was performed on an AMD Ryzen 5 5600X 6-Core CPU @ 3.7GHz, with 32GB RAM, powered by Linux. The source code was written in Python 3.8 and interpreted with PyPy 3.10. PyPy is an alternative implementation of the Python programming language, designed to be faster and more efficient in terms of memory consumption than the standard Python implementation CPython. On average, PyPy 3.10 is 4.8 times faster than CPython 3.7. Times shown are in seconds, measured as processor processing time and assuming a default value of 8~ms per page default.
	
	Specifically concerning the evaluation of SkyIR-UBS algorithm, considered datasets have been indexed using R*-tree aggregation with 4Kb page size. An associated cache with 20~\% of the corresponding R-tree’s blocks was used with every experiment.
	
	We have generated \textcolor{new}{synthetic sets of data where dimensions are independent} composed of $10$ to $1$ billion tuples for respectively $3$, $6$ and $9$ dimensions coming from the example use case.
	
	\subsection{Comparison of ranking methods}\label{ssec:comparaison_de_l_ensemble_des_solutions}
	
	To compare all methods we used the most efficient dp-idp algorithm, SkyIR-UBS.
	
\begin{figure}[htbp]
		\centering
		\resizebox{.85\linewidth}{!}{
			\begin{tikzpicture}[
				line join=bevel,
				smallskybluenode/.style={circle, fill=skyblue, draw=black, line width=0.5pt, minimum size=4pt, inner sep=0pt},
				smallcyannode/.style={circle, fill=cyan, draw=black, line width=0.5pt, minimum size=4pt, inner sep=0pt},
				smallbrightmaroonnode/.style={circle, fill=brightmaroon, draw=black, line width=0.5pt, minimum size=4pt, inner sep=0pt},
				smallSQLCodeGreennode/.style={circle, fill=SQLCodeGreen, draw=black, line width=0.5pt, minimum size=4pt, inner sep=0pt},
				smallSQLcodegraynode/.style={circle, fill=SQLcodegray, draw=black, line width=0.5pt, minimum size=4pt, inner sep=0pt},
				]
				\draw[-stealth] (0pt, 0pt) -- (280pt, 0pt) node[anchor=north west, yshift=15pt] {Cardinality};
				\draw[-stealth] (0pt, 0pt) -- (0pt, 280pt) node[anchor=south] {Response time in s};
				\foreach \x/\xtext in {
					0pt/$0$,
					56pt/$10000$,
					112pt/$20000$,
					168pt/$30000$,
					224pt/$40000$,
					280pt/$50000$} {
					\draw (\x, 2pt) -- (\x, -2pt) node[below] {\xtext\strut};
				}
				\foreach \y/\ytext in {
					0pt/$0$,
					32pt/$3000$,
					64pt/$6000$,
					96pt/$8500$,
					128pt/$11000$,
					160pt/$14000$,
					192pt/$17000$,
					224pt/$21000$} {
					\draw (2pt, \y) -- (-2pt, \y) node[left] {\ytext\strut};
				}
				\draw[skyblue, line width=2pt](0pt, 0pt) -- (0pt, 0pt) -- (0pt, 0pt) -- (1pt, 0pt) -- (1pt, 0pt) -- (3pt, 0pt) -- (6pt, 0pt) -- (11pt, 0pt) -- (28pt, 0pt) -- (56pt, 0pt) -- (112pt, 0pt) -- (280pt, 0pt);
				\draw[cyan, line width=2pt](0pt, 0pt) -- (0pt, 0pt) -- (0pt, 0pt) -- (1pt, 0pt) -- (1pt, 0pt) -- (3pt, 0pt) -- (6pt, 0pt) -- (11pt, 0pt) -- (28pt, 0pt) -- (56pt, 0pt) -- (112pt, 0pt) -- (280pt, 2pt);
				\draw[brightmaroon, line width=2pt](0pt, 0pt) -- (0pt, 0pt) -- (0pt, 0pt) -- (1pt, 0pt) -- (1pt, 0pt) -- (3pt, 0pt) -- (6pt, 0pt) -- (11pt, 0pt) -- (28pt, 0pt) -- (56pt, 0pt) -- (112pt, 0pt) -- (280pt, 1pt);
				\draw[SQLCodeGreen, line width=2pt](0pt, 0pt) -- (0pt, 0pt) -- (0pt, 0pt) -- (1pt, 0pt) -- (1pt, 0pt) -- (3pt, 0pt) -- (6pt, 0pt) -- (11pt, 0pt) -- (28pt, 0pt) -- (56pt, 1pt) -- (112pt, 8pt) -- (280pt, 108pt);
				\draw[SQLcodegray, line width=2pt](0pt, 0pt) -- (0pt, 0pt) -- (0pt, 0pt) -- (1pt, 0pt) -- (1pt, 0pt) -- (3pt, 0pt) -- (6pt, 0pt) -- (11pt, 0pt) -- (28pt, 6pt) -- (56pt, 20pt) -- (112pt, 111pt) -- (280pt, 226pt);
				\filldraw[color=black, fill=skyblue] (0pt, 0pt) circle (2pt);
				\filldraw[color=black, fill=skyblue] (0pt, 0pt) circle (2pt);
				\filldraw[color=black, fill=skyblue] (0pt, 0pt) circle (2pt);
				\filldraw[color=black, fill=skyblue] (1pt, 0pt) circle (2pt);
				\filldraw[color=black, fill=skyblue] (1pt, 0pt) circle (2pt);
				\filldraw[color=black, fill=skyblue] (3pt, 0pt) circle (2pt);
				\filldraw[color=black, fill=skyblue] (6pt, 0pt) circle (2pt);
				\filldraw[color=black, fill=skyblue] (11pt, 0pt) circle (2pt);
				\filldraw[color=black, fill=skyblue] (28pt, 0pt) circle (2pt);
				\filldraw[color=black, fill=skyblue] (56pt, 0pt) circle (2pt);
				\filldraw[color=black, fill=skyblue] (112pt, 0pt) circle (2pt);
				\filldraw[color=black, fill=skyblue] (280pt, 0pt) circle (2pt);
				\filldraw[color=black, fill=cyan] (0pt, 0pt) circle (2pt);
				\filldraw[color=black, fill=cyan] (0pt, 0pt) circle (2pt);
				\filldraw[color=black, fill=cyan] (0pt, 0pt) circle (2pt);
				\filldraw[color=black, fill=cyan] (1pt, 0pt) circle (2pt);
				\filldraw[color=black, fill=cyan] (1pt, 0pt) circle (2pt);
				\filldraw[color=black, fill=cyan] (3pt, 0pt) circle (2pt);
				\filldraw[color=black, fill=cyan] (6pt, 0pt) circle (2pt);
				\filldraw[color=black, fill=cyan] (11pt, 0pt) circle (2pt);
				\filldraw[color=black, fill=cyan] (28pt, 0pt) circle (2pt);
				\filldraw[color=black, fill=cyan] (56pt, 0pt) circle (2pt);
				\filldraw[color=black, fill=cyan] (112pt, 0pt) circle (2pt);
				\filldraw[color=black, fill=cyan] (280pt, 2pt) circle (2pt);
				\filldraw[color=black, fill=brightmaroon] (0pt, 0pt) circle (2pt);
				\filldraw[color=black, fill=brightmaroon] (0pt, 0pt) circle (2pt);
				\filldraw[color=black, fill=brightmaroon] (0pt, 0pt) circle (2pt);
				\filldraw[color=black, fill=brightmaroon] (1pt, 0pt) circle (2pt);
				\filldraw[color=black, fill=brightmaroon] (1pt, 0pt) circle (2pt);
				\filldraw[color=black, fill=brightmaroon] (3pt, 0pt) circle (2pt);
				\filldraw[color=black, fill=brightmaroon] (6pt, 0pt) circle (2pt);
				\filldraw[color=black, fill=brightmaroon] (11pt, 0pt) circle (2pt);
				\filldraw[color=black, fill=brightmaroon] (28pt, 0pt) circle (2pt);
				\filldraw[color=black, fill=brightmaroon] (56pt, 0pt) circle (2pt);
				\filldraw[color=black, fill=brightmaroon] (112pt, 0pt) circle (2pt);
				\filldraw[color=black, fill=brightmaroon] (280pt, 1pt) circle (2pt);
				\filldraw[color=black, fill=SQLCodeGreen] (0pt, 0pt) circle (2pt);
				\filldraw[color=black, fill=SQLCodeGreen] (0pt, 0pt) circle (2pt);
				\filldraw[color=black, fill=SQLCodeGreen] (0pt, 0pt) circle (2pt);
				\filldraw[color=black, fill=SQLCodeGreen] (1pt, 0pt) circle (2pt);
				\filldraw[color=black, fill=SQLCodeGreen] (1pt, 0pt) circle (2pt);
				\filldraw[color=black, fill=SQLCodeGreen] (3pt, 0pt) circle (2pt);
				\filldraw[color=black, fill=SQLCodeGreen] (6pt, 0pt) circle (2pt);
				\filldraw[color=black, fill=SQLCodeGreen] (11pt, 0pt) circle (2pt);
				\filldraw[color=black, fill=SQLCodeGreen] (28pt, 0pt) circle (2pt);
				\filldraw[color=black, fill=SQLCodeGreen] (56pt, 1pt) circle (2pt);
				\filldraw[color=black, fill=SQLCodeGreen] (112pt, 8pt) circle (2pt);
				\filldraw[color=black, fill=SQLCodeGreen] (280pt, 108pt) circle (2pt);
				\filldraw[color=black, fill=SQLcodegray] (0pt, 0pt) circle (2pt);
				\filldraw[color=black, fill=SQLcodegray] (0pt, 0pt) circle (2pt);
				\filldraw[color=black, fill=SQLcodegray] (0pt, 0pt) circle (2pt);
				\filldraw[color=black, fill=SQLcodegray] (1pt, 0pt) circle (2pt);
				\filldraw[color=black, fill=SQLcodegray] (1pt, 0pt) circle (2pt);
				\filldraw[color=black, fill=SQLcodegray] (3pt, 0pt) circle (2pt);
				\filldraw[color=black, fill=SQLcodegray] (6pt, 0pt) circle (2pt);
				\filldraw[color=black, fill=SQLcodegray] (11pt, 0pt) circle (2pt);
				\filldraw[color=black, fill=SQLcodegray] (28pt, 6pt) circle (2pt);
				\filldraw[color=black, fill=SQLcodegray] (56pt, 20pt) circle (2pt);
				\filldraw[color=black, fill=SQLcodegray] (112pt, 111pt) circle (2pt);
				\filldraw[color=black, fill=SQLcodegray] (280pt, 226pt) circle (2pt);
				\matrix [left=0cm of current bounding box.north east] at (current bounding box.north east) {
					\node [smallskybluenode, label=right:CoSky 'SQL query' with 3 attributes] {}; \\
					\node [smallcyannode, label=right:CoSky 'algorithm' with 3 attributes] {}; \\
					\node [smallbrightmaroonnode, label=right:RankSky with 3 attributes] {}; \\
					\node [smallSQLCodeGreennode, label=right:dp-idp with dominance hierarchy with 3 attributes] {}; \\
					\node [smallSQLcodegraynode, label=right:SkyIR-UBS with 3 attributes] {}; \\
				};
			\end{tikzpicture}
		}
		\caption{Response time of different solutions}\label{fig:temps_de_reponse_des_differentes_solutions}
	\end{figure}
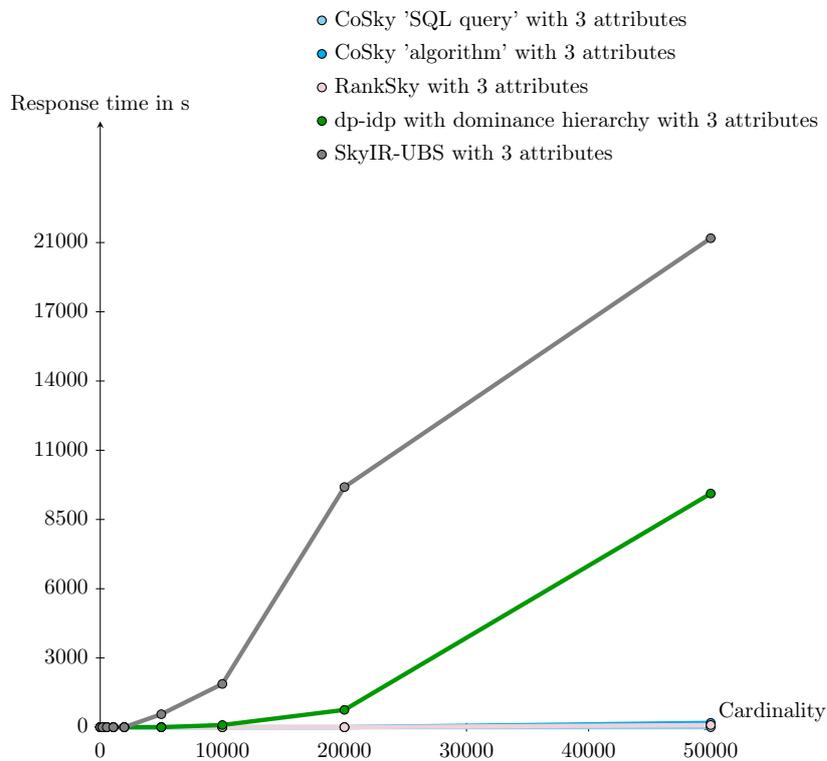

	Even though, figure~\ref{fig:temps_de_reponse_des_differentes_solutions}, portraying different solutions response times, when the dataset cardinality is up to $50000$ tuples, for $3$ dimensions, declares SkyIR-UBS as the least efficient of all methods. Our proposition, dp-idp with dominance hierarchy, explodes about twice as slow. However, its efficiency is negligible compared to RankSky and CoSky implementations. In the worst case ($50000$ tuples and $3$ dimensions), RankSky implementation, algorithmic and SQL CoSky implementations have respective response times of $1$ minute and $46$ seconds, $2$ minutes and $16$ seconds, $0.172$ seconds while our dp-idp version and SkyIR-UBS have respective response times of less than $3$ hours and more than $6$ hours!
	
	In the following evaluations, we dismiss SkyIR-UBS and dp-idp with dominance hierarchy, less efficient than RankSky and CoSky implementations, in order to study higher cardinalities and higher dimensions.
	
	\subsection{RankSky and CoSky implementations}\label{ssec:comparaison_des_implementations_de_CoSky}
	
\begin{figure}[htbp]
		\centering
		\resizebox{.45\linewidth}{!}{
			\begin{tikzpicture}[
				line join=bevel,
				smallskybluenode/.style={circle, fill=skyblue, draw=black, line width=0.5pt, minimum size=4pt, inner sep=0pt},
				smallcyannode/.style={circle, fill=cyan, draw=black, line width=0.5pt, minimum size=4pt, inner sep=0pt},
				smallbrightmaroonnode/.style={circle, fill=brightmaroon, draw=black, line width=0.5pt, minimum size=4pt, inner sep=0pt},
				]
				\draw[-stealth] (0pt, 0pt) -- (280pt, 0pt) node[anchor=north west, yshift=15pt] {Cardinality};
				\draw[-stealth] (0pt, 0pt) -- (0pt, 280pt) node[anchor=south, font=\large] {Response time in s};
				\foreach \x/\xtext in {
					0pt/$0$, 
					56pt/$40000$, 
					112pt/$80000$, 
					168pt/$120000$, 
					224pt/$160000$, 
					280pt/$200000$} {
					\draw (\x, 2pt) -- (\x, -2pt) node[below, font=\small] {\xtext\strut};
				}
				\foreach \y/\ytext in {
					0pt/$0$, 
					220pt/$400$, 
					246pt/$800$, 
					261pt/$1200$, 
					271pt/$1600$, 
					280pt/$2000$} {
					\draw (2pt, \y) -- (-2pt, \y) node[left, font=\small] {\ytext\strut};
				}
				\draw[skyblue, line width=2pt](0pt, 0pt) -- (0pt, 0pt) -- (0pt, 0pt) -- (0pt, 0pt) -- (0pt, 0pt) -- (1pt, 0pt) -- (1pt, 0pt) -- (3pt, 0pt) -- (7pt, 1pt) -- (14pt, 1pt) -- (28pt, 3pt) -- (70pt, 6pt) -- (140pt, 19pt) -- (280pt, 30pt);
				\draw[cyan, line width=2pt](0pt, 1pt) -- (0pt, 0pt) -- (0pt, 1pt) -- (0pt, 1pt) -- (0pt, 1pt) -- (1pt, 1pt) -- (1pt, 3pt) -- (3pt, 7pt) -- (7pt, 23pt) -- (14pt, 52pt) -- (28pt, 111pt) -- (70pt, 174pt) -- (140pt, 239pt) -- (280pt, 285pt);
				\draw[brightmaroon, line width=2pt](0pt, 0pt) -- (0pt, 0pt) -- (0pt, 0pt) -- (0pt, 0pt) -- (0pt, 1pt) -- (1pt, 1pt) -- (1pt, 3pt) -- (3pt, 5pt) -- (7pt, 23pt) -- (14pt, 51pt) -- (28pt, 110pt) -- (70pt, 172pt) -- (140pt, 235pt) -- (280pt, 280pt);
				\filldraw[color=black, fill=skyblue] (0pt, 0pt) circle (2pt);
				\filldraw[color=black, fill=skyblue] (0pt, 0pt) circle (2pt);
				\filldraw[color=black, fill=skyblue] (0pt, 0pt) circle (2pt);
				\filldraw[color=black, fill=skyblue] (0pt, 0pt) circle (2pt);
				\filldraw[color=black, fill=skyblue] (0pt, 0pt) circle (2pt);
				\filldraw[color=black, fill=skyblue] (1pt, 0pt) circle (2pt);
				\filldraw[color=black, fill=skyblue] (1pt, 0pt) circle (2pt);
				\filldraw[color=black, fill=skyblue] (3pt, 0pt) circle (2pt);
				\filldraw[color=black, fill=skyblue] (7pt, 1pt) circle (2pt);
				\filldraw[color=black, fill=skyblue] (14pt, 1pt) circle (2pt);
				\filldraw[color=black, fill=skyblue] (28pt, 3pt) circle (2pt);
				\filldraw[color=black, fill=skyblue] (70pt, 6pt) circle (2pt);
				\filldraw[color=black, fill=skyblue] (140pt, 19pt) circle (2pt);
				\filldraw[color=black, fill=skyblue] (280pt, 30pt) circle (2pt);
				\filldraw[color=black, fill=cyan] (0pt, 1pt) circle (2pt);
				\filldraw[color=black, fill=cyan] (0pt, 0pt) circle (2pt);
				\filldraw[color=black, fill=cyan] (0pt, 1pt) circle (2pt);
				\filldraw[color=black, fill=cyan] (0pt, 1pt) circle (2pt);
				\filldraw[color=black, fill=cyan] (0pt, 1pt) circle (2pt);
				\filldraw[color=black, fill=cyan] (1pt, 1pt) circle (2pt);
				\filldraw[color=black, fill=cyan] (1pt, 3pt) circle (2pt);
				\filldraw[color=black, fill=cyan] (3pt, 7pt) circle (2pt);
				\filldraw[color=black, fill=cyan] (7pt, 23pt) circle (2pt);
				\filldraw[color=black, fill=cyan] (14pt, 52pt) circle (2pt);
				\filldraw[color=black, fill=cyan] (28pt, 111pt) circle (2pt);
				\filldraw[color=black, fill=cyan] (70pt, 174pt) circle (2pt);
				\filldraw[color=black, fill=cyan] (140pt, 239pt) circle (2pt);
				\filldraw[color=black, fill=cyan] (280pt, 285pt) circle (2pt);
				\filldraw[color=black, fill=brightmaroon] (0pt, 0pt) circle (2pt);
				\filldraw[color=black, fill=brightmaroon] (0pt, 0pt) circle (2pt);
				\filldraw[color=black, fill=brightmaroon] (0pt, 0pt) circle (2pt);
				\filldraw[color=black, fill=brightmaroon] (0pt, 0pt) circle (2pt);
				\filldraw[color=black, fill=brightmaroon] (0pt, 1pt) circle (2pt);
				\filldraw[color=black, fill=brightmaroon] (1pt, 1pt) circle (2pt);
				\filldraw[color=black, fill=brightmaroon] (1pt, 3pt) circle (2pt);
				\filldraw[color=black, fill=brightmaroon] (3pt, 5pt) circle (2pt);
				\filldraw[color=black, fill=brightmaroon] (7pt, 23pt) circle (2pt);
				\filldraw[color=black, fill=brightmaroon] (14pt, 51pt) circle (2pt);
				\filldraw[color=black, fill=brightmaroon] (28pt, 110pt) circle (2pt);
				\filldraw[color=black, fill=brightmaroon] (70pt, 172pt) circle (2pt);
				\filldraw[color=black, fill=brightmaroon] (140pt, 235pt) circle (2pt);
				\filldraw[color=black, fill=brightmaroon] (280pt, 280pt) circle (2pt);
				\matrix [draw=none, fill=white, fill opacity=0.7, text opacity=1,
				above right=0.5cm and 0cm of current bounding box.north west,
				nodes={font=\small, anchor=west}] {
					\node [smallskybluenode, label=right:CoSky 'SQL query' with 3 attributes] {}; \\
					\node [smallcyannode, label=right:CoSky 'algorithm' with 3 attributes] {}; \\
					\node [smallbrightmaroonnode, label=right:RankSky with 3 attributes] {}; \\
				};
			\end{tikzpicture}
		}
		\resizebox{.45\linewidth}{!}{
			\begin{tikzpicture}[
				line join=bevel,
				smallskybluenode/.style={circle, fill=skyblue, draw=black, line width=0.5pt, minimum size=4pt, inner sep=0pt},
				smallcyannode/.style={circle, fill=cyan, draw=black, line width=0.5pt, minimum size=4pt, inner sep=0pt},
				smallbrightmaroonnode/.style={circle, fill=brightmaroon, draw=black, line width=0.5pt, minimum size=4pt, inner sep=0pt},
				]
				\draw[-stealth] (0pt, 0pt) -- (280pt, 0pt) node[anchor=north west, yshift=15pt] {Cardinality};
				\draw[-stealth] (0pt, 0pt) -- (0pt, 280pt) node[anchor=south, font=\large] {Response time in s};
				\foreach \x/\xtext in {
					0pt/$0$, 
					56pt/$40000$, 
					112pt/$80000$, 
					168pt/$120000$, 
					224pt/$160000$, 
					280pt/$200000$} {
					\draw (\x, 2pt) -- (\x, -2pt) node[below, font=\small] {\xtext\strut};
				}
				\foreach \y/\ytext in {
					0pt/$0$, 
					223pt/$600$, 
					247pt/$1200$, 
					262pt/$1800$, 
					272pt/$2400$, 
					280pt/$3000$} {
					\draw (2pt, \y) -- (-2pt, \y) node[left, font=\small] {\ytext\strut};
				}
				\draw[skyblue, line width=2pt](0pt, 0pt) -- (0pt, 0pt) -- (0pt, 0pt) -- (0pt, 0pt) -- (0pt, 0pt) -- (1pt, 0pt) -- (1pt, 1pt) -- (3pt, 3pt) -- (7pt, 8pt) -- (14pt, 17pt) -- (28pt, 36pt) -- (70pt, 74pt) -- (140pt, 104pt) -- (280pt, 143pt);
				\draw[cyan, line width=2pt](0pt, 0pt) -- (0pt, 0pt) -- (0pt, 0pt) -- (0pt, 0pt) -- (0pt, 2pt) -- (1pt, 1pt) -- (1pt, 2pt) -- (3pt, 5pt) -- (7pt, 27pt) -- (14pt, 61pt) -- (28pt, 108pt) -- (70pt, 169pt) -- (140pt, 225pt) -- (280pt, 285pt);
				\draw[brightmaroon, line width=2pt](0pt, 0pt) -- (0pt, 0pt) -- (0pt, 0pt) -- (0pt, 0pt) -- (0pt, 1pt) -- (1pt, 2pt) -- (1pt, 5pt) -- (3pt, 16pt) -- (7pt, 33pt) -- (14pt, 63pt) -- (28pt, 109pt) -- (70pt, 167pt) -- (140pt, 222pt) -- (280pt, 275pt);
				\filldraw[color=black, fill=skyblue] (0pt, 0pt) circle (2pt);
				\filldraw[color=black, fill=skyblue] (0pt, 0pt) circle (2pt);
				\filldraw[color=black, fill=skyblue] (0pt, 0pt) circle (2pt);
				\filldraw[color=black, fill=skyblue] (0pt, 0pt) circle (2pt);
				\filldraw[color=black, fill=skyblue] (0pt, 0pt) circle (2pt);
				\filldraw[color=black, fill=skyblue] (1pt, 0pt) circle (2pt);
				\filldraw[color=black, fill=skyblue] (1pt, 1pt) circle (2pt);
				\filldraw[color=black, fill=skyblue] (3pt, 3pt) circle (2pt);
				\filldraw[color=black, fill=skyblue] (7pt, 8pt) circle (2pt);
				\filldraw[color=black, fill=skyblue] (14pt, 17pt) circle (2pt);
				\filldraw[color=black, fill=skyblue] (28pt, 36pt) circle (2pt);
				\filldraw[color=black, fill=skyblue] (70pt, 74pt) circle (2pt);
				\filldraw[color=black, fill=skyblue] (140pt, 104pt) circle (2pt);
				\filldraw[color=black, fill=skyblue] (280pt, 143pt) circle (2pt);
				\filldraw[color=black, fill=cyan] (0pt, 0pt) circle (2pt);
				\filldraw[color=black, fill=cyan] (0pt, 0pt) circle (2pt);
				\filldraw[color=black, fill=cyan] (0pt, 0pt) circle (2pt);
				\filldraw[color=black, fill=cyan] (0pt, 0pt) circle (2pt);
				\filldraw[color=black, fill=cyan] (0pt, 2pt) circle (2pt);
				\filldraw[color=black, fill=cyan] (1pt, 1pt) circle (2pt);
				\filldraw[color=black, fill=cyan] (1pt, 2pt) circle (2pt);
				\filldraw[color=black, fill=cyan] (3pt, 5pt) circle (2pt);
				\filldraw[color=black, fill=cyan] (7pt, 27pt) circle (2pt);
				\filldraw[color=black, fill=cyan] (14pt, 61pt) circle (2pt);
				\filldraw[color=black, fill=cyan] (28pt, 108pt) circle (2pt);
				\filldraw[color=black, fill=cyan] (70pt, 169pt) circle (2pt);
				\filldraw[color=black, fill=cyan] (140pt, 225pt) circle (2pt);
				\filldraw[color=black, fill=cyan] (280pt, 285pt) circle (2pt);
				\filldraw[color=black, fill=brightmaroon] (0pt, 0pt) circle (2pt);
				\filldraw[color=black, fill=brightmaroon] (0pt, 0pt) circle (2pt);
				\filldraw[color=black, fill=brightmaroon] (0pt, 0pt) circle (2pt);
				\filldraw[color=black, fill=brightmaroon] (0pt, 0pt) circle (2pt);
				\filldraw[color=black, fill=brightmaroon] (0pt, 1pt) circle (2pt);
				\filldraw[color=black, fill=brightmaroon] (1pt, 2pt) circle (2pt);
				\filldraw[color=black, fill=brightmaroon] (1pt, 5pt) circle (2pt);
				\filldraw[color=black, fill=brightmaroon] (3pt, 16pt) circle (2pt);
				\filldraw[color=black, fill=brightmaroon] (7pt, 33pt) circle (2pt);
				\filldraw[color=black, fill=brightmaroon] (14pt, 63pt) circle (2pt);
				\filldraw[color=black, fill=brightmaroon] (28pt, 109pt) circle (2pt);
				\filldraw[color=black, fill=brightmaroon] (70pt, 167pt) circle (2pt);
				\filldraw[color=black, fill=brightmaroon] (140pt, 222pt) circle (2pt);
				\filldraw[color=black, fill=brightmaroon] (280pt, 275pt) circle (2pt);
				\matrix [draw=none, fill=white, fill opacity=0.7, text opacity=1,
				above right=0.5cm and 0cm of current bounding box.north west,
				nodes={font=\small, anchor=west}] {
					\node [smallskybluenode, label=right:CoSky 'SQL query' with 6 attributes] {}; \\
					\node [smallcyannode, label=right:CoSky 'algorithm' with 6 attributes] {}; \\
					\node [smallbrightmaroonnode, label=right:RankSky with 6 attributes] {}; \\
				};
			\end{tikzpicture}
		}
		\resizebox{.45\linewidth}{!}{
			\begin{tikzpicture}[
				line join=bevel,
				smallskybluenode/.style={circle, fill=skyblue, draw=black, line width=0.5pt, minimum size=4pt, inner sep=0pt},
				smallcyannode/.style={circle, fill=cyan, draw=black, line width=0.5pt, minimum size=4pt, inner sep=0pt},
				smallbrightmaroonnode/.style={circle, fill=brightmaroon, draw=black, line width=0.5pt, minimum size=4pt, inner sep=0pt},
				]
				\draw[-stealth] (0pt, 0pt) -- (280pt, 0pt) node[anchor=north west, yshift=15pt] {Cardinality};
				\draw[-stealth] (0pt, 0pt) -- (0pt, 280pt) node[anchor=south, font=\large] {Response time in s};
				\foreach \x/\xtext in {
					0pt/$0$, 
					56pt/$40000$, 
					112pt/$80000$, 
					168pt/$120000$, 
					224pt/$160000$, 
					280pt/$200000$} {
					\draw (\x, 2pt) -- (\x, -2pt) node[below, font=\small] {\xtext\strut};
				}
				\foreach \y/\ytext in {
					0pt/$0$, 
					227pt/$1000$, 
					249pt/$2000$, 
					263pt/$3000$, 
					272pt/$4000$, 
					280pt/$5000$} {
					\draw (2pt, \y) -- (-2pt, \y) node[left, font=\small] {\ytext\strut};
				}
				\draw[skyblue, line width=2pt](0pt, 0pt) -- (0pt, 0pt) -- (0pt, 0pt) -- (0pt, 0pt) -- (0pt, 0pt) -- (1pt, 1pt) -- (1pt, 2pt) -- (3pt, 5pt) -- (7pt, 20pt) -- (14pt, 41pt) -- (28pt, 74pt) -- (70pt, 129pt) -- (140pt, 163pt) -- (280pt, 203pt);
				\draw[cyan, line width=2pt](0pt, 0pt) -- (0pt, 0pt) -- (0pt, 0pt) -- (0pt, 0pt) -- (0pt, 0pt) -- (1pt, 1pt) -- (1pt, 2pt) -- (3pt, 7pt) -- (7pt, 34pt) -- (14pt, 68pt) -- (28pt, 121pt) -- (70pt, 169pt) -- (140pt, 236pt) -- (280pt, 283pt);
				\draw[brightmaroon, line width=2pt](0pt, 0pt) -- (0pt, 0pt) -- (0pt, 0pt) -- (0pt, 1pt) -- (0pt, 2pt) -- (1pt, 8pt) -- (1pt, 25pt) -- (3pt, 45pt) -- (7pt, 84pt) -- (14pt, 112pt) -- (28pt, 153pt) -- (70pt, 193pt) -- (140pt, 241pt) -- (280pt, 279pt);
				\filldraw[color=black, fill=skyblue] (0pt, 0pt) circle (2pt);
				\filldraw[color=black, fill=skyblue] (0pt, 0pt) circle (2pt);
				\filldraw[color=black, fill=skyblue] (0pt, 0pt) circle (2pt);
				\filldraw[color=black, fill=skyblue] (0pt, 0pt) circle (2pt);
				\filldraw[color=black, fill=skyblue] (0pt, 0pt) circle (2pt);
				\filldraw[color=black, fill=skyblue] (1pt, 1pt) circle (2pt);
				\filldraw[color=black, fill=skyblue] (1pt, 2pt) circle (2pt);
				\filldraw[color=black, fill=skyblue] (3pt, 5pt) circle (2pt);
				\filldraw[color=black, fill=skyblue] (7pt, 20pt) circle (2pt);
				\filldraw[color=black, fill=skyblue] (14pt, 41pt) circle (2pt);
				\filldraw[color=black, fill=skyblue] (28pt, 74pt) circle (2pt);
				\filldraw[color=black, fill=skyblue] (70pt, 129pt) circle (2pt);
				\filldraw[color=black, fill=skyblue] (140pt, 163pt) circle (2pt);
				\filldraw[color=black, fill=skyblue] (280pt, 203pt) circle (2pt);
				\filldraw[color=black, fill=cyan] (0pt, 0pt) circle (2pt);
				\filldraw[color=black, fill=cyan] (0pt, 0pt) circle (2pt);
				\filldraw[color=black, fill=cyan] (0pt, 0pt) circle (2pt);
				\filldraw[color=black, fill=cyan] (0pt, 0pt) circle (2pt);
				\filldraw[color=black, fill=cyan] (0pt, 0pt) circle (2pt);
				\filldraw[color=black, fill=cyan] (1pt, 1pt) circle (2pt);
				\filldraw[color=black, fill=cyan] (1pt, 2pt) circle (2pt);
				\filldraw[color=black, fill=cyan] (3pt, 7pt) circle (2pt);
				\filldraw[color=black, fill=cyan] (7pt, 34pt) circle (2pt);
				\filldraw[color=black, fill=cyan] (14pt, 68pt) circle (2pt);
				\filldraw[color=black, fill=cyan] (28pt, 121pt) circle (2pt);
				\filldraw[color=black, fill=cyan] (70pt, 169pt) circle (2pt);
				\filldraw[color=black, fill=cyan] (140pt, 236pt) circle (2pt);
				\filldraw[color=black, fill=cyan] (280pt, 283pt) circle (2pt);
				\filldraw[color=black, fill=brightmaroon] (0pt, 0pt) circle (2pt);
				\filldraw[color=black, fill=brightmaroon] (0pt, 0pt) circle (2pt);
				\filldraw[color=black, fill=brightmaroon] (0pt, 0pt) circle (2pt);
				\filldraw[color=black, fill=brightmaroon] (0pt, 1pt) circle (2pt);
				\filldraw[color=black, fill=brightmaroon] (0pt, 2pt) circle (2pt);
				\filldraw[color=black, fill=brightmaroon] (1pt, 8pt) circle (2pt);
				\filldraw[color=black, fill=brightmaroon] (1pt, 25pt) circle (2pt);
				\filldraw[color=black, fill=brightmaroon] (3pt, 45pt) circle (2pt);
				\filldraw[color=black, fill=brightmaroon] (7pt, 84pt) circle (2pt);
				\filldraw[color=black, fill=brightmaroon] (14pt, 112pt) circle (2pt);
				\filldraw[color=black, fill=brightmaroon] (28pt, 153pt) circle (2pt);
				\filldraw[color=black, fill=brightmaroon] (70pt, 193pt) circle (2pt);
				\filldraw[color=black, fill=brightmaroon] (140pt, 241pt) circle (2pt);
				\filldraw[color=black, fill=brightmaroon] (280pt, 279pt) circle (2pt);
				\matrix [draw=none, fill=white, fill opacity=0.7, text opacity=1,
				above right=0.5cm and 0cm of current bounding box.north west,
				nodes={font=\small, anchor=west}] {
					\node [smallskybluenode, label=right:CoSky 'SQL query' with 9 attributes] {}; \\
					\node [smallcyannode, label=right:CoSky 'algorithm' with 9 attributes] {}; \\
					\node [smallbrightmaroonnode, label=right:RankSky with 9 attributes] {}; \\
				};
			\end{tikzpicture}
		}
		\caption{CoSky response time Sql query vs Algorithm (3, 6 and 9 attributes)}\label{fig:temps_de_reponse_des_requetes_CoSky}
	\end{figure}
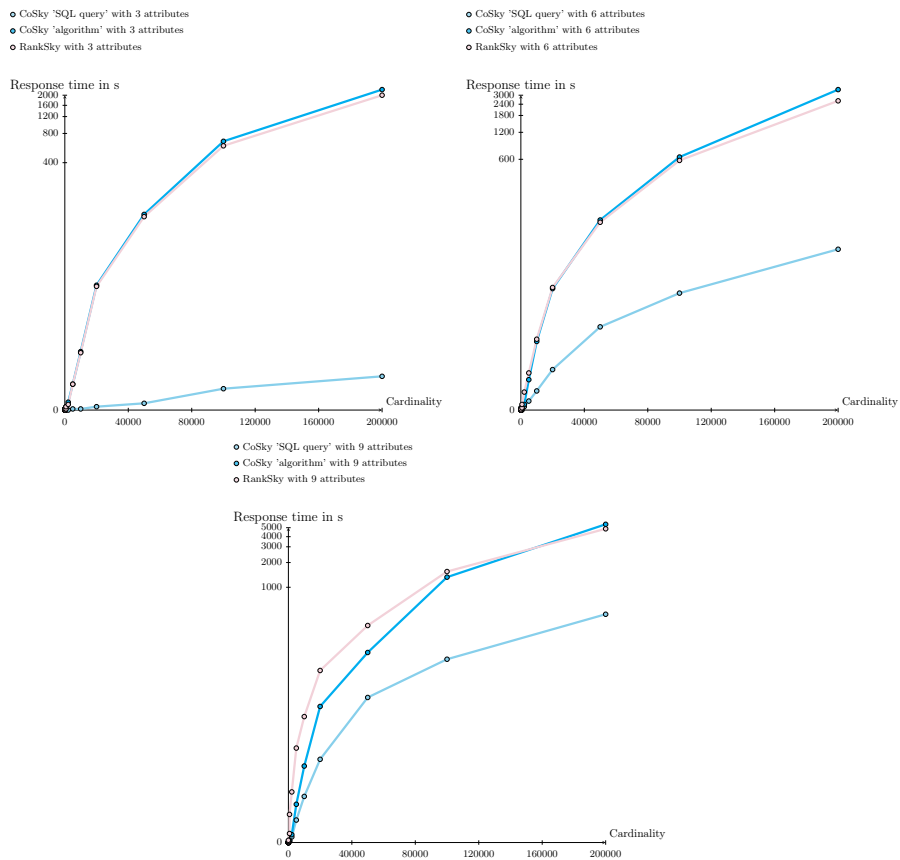

	To compare RankSky and CoSky's implementations, we consider figure~\ref{fig:temps_de_reponse_des_requetes_CoSky} with logarithmic scale representation for ordinate axis that, in our humble opinion, provides a more relevant visualization. Evaluation are made for cardinalities up to $200000$ tuples with, respectively, $3$, $6$ and $9$ dimensions. In the worst case, response times are approximately $34$ minutes, $44$ minutes and $1$ hour $20$ minutes respectively, for RankSky, while they are approximately $38$ minutes, $57$ minutes and $1$ hour $31$ minutes respectively, for algorithmic CoSky while they are approximately $1$ second, $58$ seconds and $7$ minutes and $56$ seconds respectively, for SQL CoSky. Even though, RankSky and both CoSky propositions are particularly efficient, notably in regards to existing methods. RankSky and algorithmic CoSky are quite similar. The DBMS integrated CoSky method is significantly faster.
	
	\subsection{SQL CoSky}\label{ssec:comparaison_de_l_implementation_sql_de_CoSky}
	
\begin{figure}[htbp]
		\centering
		\resizebox{.85\linewidth}{!}{
			\begin{tikzpicture}[line join=bevel,
				bigbrightmaroonnode/.style={shape=circle, fill=brightmaroon, draw=black, line width=1pt},
				bigcyannode/.style={shape=circle, fill=cyan, draw=black, line width=1pt},
				bigskybluenode/.style={shape=circle, fill=skyblue, draw=black, line width=1pt}]
				\draw[-stealth] (0pt, 0pt) -- (280pt, 0pt) node[anchor=north west, yshift=15pt] {Cardinality};
				\draw[-stealth] (0pt, 0pt) -- (0pt, 280pt) node[anchor=south] {Response time in s};
				\foreach \x/\xtext in {
					0pt/$0$,
					56pt/$400000$,
					112pt/$800000$,
					168pt/$1200000$,
					224pt/$1600000$,
					280pt/$2000000$} {
					\draw (\x, 2pt) -- (\x, -2pt) node[below] {\xtext\strut};
				}
				\foreach \y/\ytext in {
					0pt/$0$,
					44pt/$4000$,
					89pt/$8000$,
					134pt/$12000$,
					179pt/$16000$,
					224pt/$20000$} {
					\draw (2pt, \y) -- (-2pt, \y) node[left] {\ytext\strut};
				}
				\draw[skyblue, line width=2pt] (0pt, 0pt) -- (0pt, 0pt) -- (0pt, 0pt) -- (0pt, 0pt) -- (0pt, 0pt) -- (0pt, 0pt) -- (0pt, 0pt) -- (0pt, 0pt) -- (1pt, 0pt) -- (1pt, 0pt) -- (3pt, 0pt) -- (7pt, 0pt) -- (14pt, 0pt) -- (28pt, 0pt) -- (70pt, 0pt) -- (140pt, 0pt) -- (280pt, 0pt);
				\filldraw[color=black, fill=skyblue] (0pt, 0pt) circle (2pt);
				\filldraw[color=black, fill=skyblue] (0pt, 0pt) circle (2pt);
				\filldraw[color=black, fill=skyblue] (0pt, 0pt) circle (2pt);
				\filldraw[color=black, fill=skyblue] (0pt, 0pt) circle (2pt);
				\filldraw[color=black, fill=skyblue] (0pt, 0pt) circle (2pt);
				\filldraw[color=black, fill=skyblue] (0pt, 0pt) circle (2pt);
				\filldraw[color=black, fill=skyblue] (0pt, 0pt) circle (2pt);
				\filldraw[color=black, fill=skyblue] (0pt, 0pt) circle (2pt);
				\filldraw[color=black, fill=skyblue] (1pt, 0pt) circle (2pt);
				\filldraw[color=black, fill=skyblue] (1pt, 0pt) circle (2pt);
				\filldraw[color=black, fill=skyblue] (3pt, 0pt) circle (2pt);
				\filldraw[color=black, fill=skyblue] (7pt, 0pt) circle (2pt);
				\filldraw[color=black, fill=skyblue] (14pt, 0pt) circle (2pt);
				\filldraw[color=black, fill=skyblue] (28pt, 0pt) circle (2pt);
				\filldraw[color=black, fill=skyblue] (70pt, 0pt) circle (2pt);
				\filldraw[color=black, fill=skyblue] (140pt, 0pt) circle (2pt);
				\filldraw[color=black, fill=skyblue] (280pt, 0pt) circle (2pt);
				
				\draw[cyan, line width=2pt] (0pt, 0pt) -- (0pt, 0pt) -- (0pt, 0pt) -- (0pt, 0pt) -- (0pt, 0pt) -- (0pt, 0pt) -- (0pt, 0pt) -- (0pt, 0pt) -- (1pt, 0pt) -- (1pt, 0pt) -- (3pt, 0pt) -- (7pt, 0pt) -- (14pt, 0pt) -- (28pt, 1pt) -- (70pt, 2pt) -- (140pt, 5pt) -- (280pt, 12pt);
				\filldraw[color=black, fill=cyan] (0pt, 0pt) circle (2pt);
				\filldraw[color=black, fill=cyan] (0pt, 0pt) circle (2pt);
				\filldraw[color=black, fill=cyan] (0pt, 0pt) circle (2pt);
				\filldraw[color=black, fill=cyan] (0pt, 0pt) circle (2pt);
				\filldraw[color=black, fill=cyan] (0pt, 0pt) circle (2pt);
				\filldraw[color=black, fill=cyan] (0pt, 0pt) circle (2pt);
				\filldraw[color=black, fill=cyan] (0pt, 0pt) circle (2pt);
				\filldraw[color=black, fill=cyan] (0pt, 0pt) circle (2pt);
				\filldraw[color=black, fill=cyan] (1pt, 0pt) circle (2pt);
				\filldraw[color=black, fill=cyan] (1pt, 0pt) circle (2pt);
				\filldraw[color=black, fill=cyan] (3pt, 0pt) circle (2pt);
				\filldraw[color=black, fill=cyan] (7pt, 0pt) circle (2pt);
				\filldraw[color=black, fill=cyan] (14pt, 0pt) circle (2pt);
				\filldraw[color=black, fill=cyan] (28pt, 1pt) circle (2pt);
				\filldraw[color=black, fill=cyan] (70pt, 2pt) circle (2pt);
				\filldraw[color=black, fill=cyan] (140pt, 5pt) circle (2pt);
				\filldraw[color=black, fill=cyan] (280pt, 12pt) circle (2pt);
				
				\draw[brightmaroon, line width=2pt] (0pt, 0pt) -- (0pt, 0pt) -- (0pt, 0pt) -- (0pt, 0pt) -- (0pt, 0pt) -- (0pt, 0pt) -- (0pt, 0pt) -- (0pt, 0pt) -- (1pt, 0pt) -- (1pt, 0pt) -- (3pt, 0pt) -- (7pt, 1pt) -- (14pt, 2pt) -- (28pt, 5pt) -- (70pt, 21pt) -- (140pt, 62pt) -- (280pt, 179pt);
				\filldraw[color=black, fill=brightmaroon] (0pt, 0pt) circle (2pt);
				\filldraw[color=black, fill=brightmaroon] (0pt, 0pt) circle (2pt);
				\filldraw[color=black, fill=brightmaroon] (0pt, 0pt) circle (2pt);
				\filldraw[color=black, fill=brightmaroon] (0pt, 0pt) circle (2pt);
				\filldraw[color=black, fill=brightmaroon] (0pt, 0pt) circle (2pt);
				\filldraw[color=black, fill=brightmaroon] (0pt, 0pt) circle (2pt);
				\filldraw[color=black, fill=brightmaroon] (0pt, 0pt) circle (2pt);
				\filldraw[color=black, fill=brightmaroon] (0pt, 0pt) circle (2pt);
				\filldraw[color=black, fill=brightmaroon] (1pt, 0pt) circle (2pt);
				\filldraw[color=black, fill=brightmaroon] (1pt, 0pt) circle (2pt);
				\filldraw[color=black, fill=brightmaroon] (3pt, 0pt) circle (2pt);
				\filldraw[color=black, fill=brightmaroon] (7pt, 1pt) circle (2pt);
				\filldraw[color=black, fill=brightmaroon] (14pt, 2pt) circle (2pt);
				\filldraw[color=black, fill=brightmaroon] (28pt, 5pt) circle (2pt);
				\filldraw[color=black, fill=brightmaroon] (70pt, 21pt) circle (2pt);
				\filldraw[color=black, fill=brightmaroon] (140pt, 62pt) circle (2pt);
				\filldraw[color=black, fill=brightmaroon] (280pt, 179pt) circle (2pt);
				
				\matrix [below left] at (current bounding box.north east) {
					\node [bigskybluenode, label=right:CoSky 'SQL query' with 3 attributes] {}; \\
					\node [bigcyannode, label=right:CoSky 'SQL query' with 6 attributes] {}; \\
					\node [bigbrightmaroonnode, label=right:CoSky 'SQL query' with 9 attributes] {}; \\
				};
			\end{tikzpicture}
		}
		\caption{CoSky response time for SQL queries}\label{fig:temps_de_reponse_de_CoSky_en_sql}
	\end{figure}
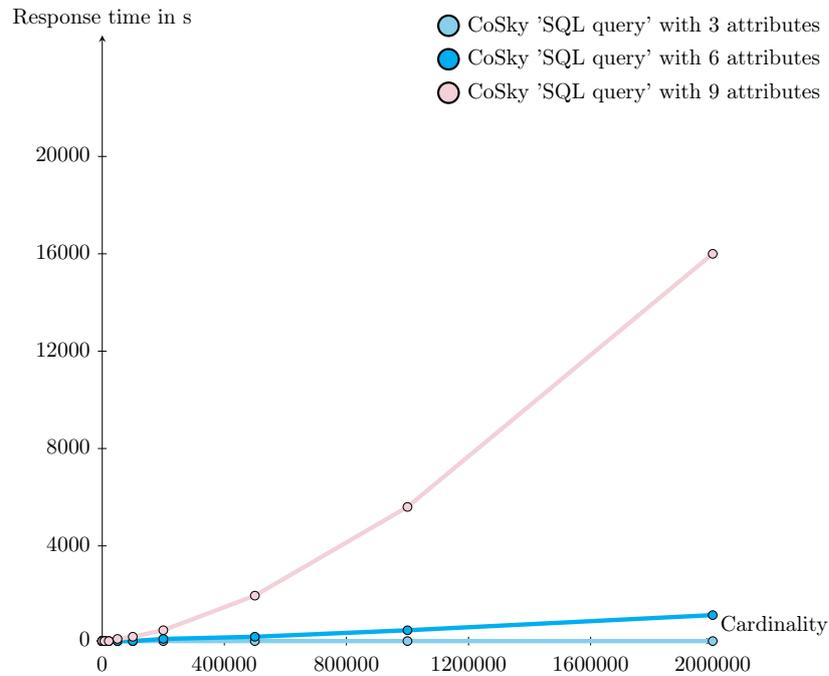

	To compare SQL CoSky implementations, we consider figure~\ref{fig:temps_de_reponse_de_CoSky_en_sql}. Evaluation are made for cardinalities up to $2~000~000$ tuples with, respectively, $3$, $6$ and $9$ dimensions. In the worst case, response times are approximately about $11.08$ seconds for $3$ dimensions, about $18$ minutes for $6$ dimensions and about $4$ hours $26$ minutes for $9$ dimensions. \textcolor{new}{As expected, our SQL CoSky implementation is sensitive to an increase in the number of dimensions, as is the case with any kind of computation in a Skyline context.} This problem is known to be difficult for high dimensions even with a RAM model (\cite{chanHighDimensionalSkylines2006}). 
	
	\subsection{CoSky with 3 dimensions}\label{ssec:evaluation_de_l_implementation_sql_de_CoSky_avec_3_dimensions}
	
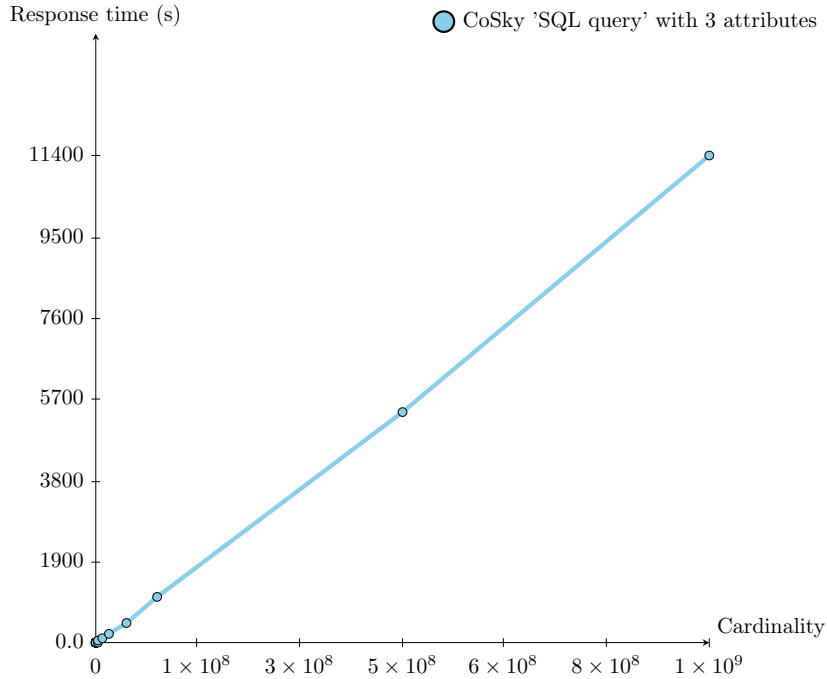
\begin{figure}[htbp]
		\centering
		\resizebox{.85\linewidth}{!}{
			\begin{tikzpicture}[
				line join=bevel,
				bigcyannode/.style={shape=circle, fill=cyan, draw=black, line width=1pt},
				bigskybluenode/.style={shape=circle, fill=skyblue, draw=black, line width=1pt}
				]
				\draw[-stealth] (0pt, 0pt) -- (280pt, 0pt) node[anchor=north west, yshift=15pt] {Cardinality};
				\draw[-stealth] (0pt, 0pt) -- (0pt, 280pt) node[anchor=south] {Response time (s)};
				
				\foreach \x/\xtext in {
					0pt/$0$,
					46pt/$1 \times 10^{8}$,
					93pt/$3 \times 10^{8}$,
					140pt/$5 \times 10^{8}$,
					186pt/$6 \times 10^{8}$,
					233pt/$8 \times 10^{8}$,
					280pt/$1 \times 10^{9}$} {
					\draw (\x, 2pt) -- (\x, -2pt) node[below] {\xtext\strut};
				}
				
				\foreach \y/\ytext in {
					0pt/$0.0$,
					37pt/$1900$,
					74pt/$3800$,
					112pt/$5700$,
					149pt/$7600$,
					186pt/$9500$,
					224pt/$11400$} {
					\draw (2pt, \y) -- (-2pt, \y) node[left] {\ytext\strut};
				}
				
				\draw[skyblue, line width=2pt] (0pt, 0pt) -- (0pt, 0pt) -- (0pt, 0pt) -- (0pt, 0pt) -- (0pt, 0pt) -- (0pt, 0pt) -- (0pt, 0pt) -- (0pt, 0pt) -- (0pt, 0pt) -- (0pt, 0pt) -- (0pt, 0pt) -- (0pt, 0pt) -- (0pt, 0pt) -- (0pt, 0pt) -- (0pt, 0pt) -- (0pt, 0pt) -- (1pt, 0pt) -- (1pt, 1pt) -- (3pt, 2pt) -- (6pt, 4pt) -- (14pt, 9pt) -- (28pt, 21pt) -- (140pt, 106pt) -- (280pt, 224pt);
				\filldraw[color=black, fill=skyblue] (0pt, 0pt) circle (2pt);
				\filldraw[color=black, fill=skyblue] (0pt, 0pt) circle (2pt);
				\filldraw[color=black, fill=skyblue] (0pt, 0pt) circle (2pt);
				\filldraw[color=black, fill=skyblue] (0pt, 0pt) circle (2pt);
				\filldraw[color=black, fill=skyblue] (0pt, 0pt) circle (2pt);
				\filldraw[color=black, fill=skyblue] (0pt, 0pt) circle (2pt);
				\filldraw[color=black, fill=skyblue] (0pt, 0pt) circle (2pt);
				\filldraw[color=black, fill=skyblue] (0pt, 0pt) circle (2pt);
				\filldraw[color=black, fill=skyblue] (0pt, 0pt) circle (2pt);
				\filldraw[color=black, fill=skyblue] (0pt, 0pt) circle (2pt);
				\filldraw[color=black, fill=skyblue] (0pt, 0pt) circle (2pt);
				\filldraw[color=black, fill=skyblue] (0pt, 0pt) circle (2pt);
				\filldraw[color=black, fill=skyblue] (0pt, 0pt) circle (2pt);
				\filldraw[color=black, fill=skyblue] (0pt, 0pt) circle (2pt);
				\filldraw[color=black, fill=skyblue] (0pt, 0pt) circle (2pt);
				\filldraw[color=black, fill=skyblue] (0pt, 0pt) circle (2pt);
				\filldraw[color=black, fill=skyblue] (1pt, 0pt) circle (2pt);
				\filldraw[color=black, fill=skyblue] (1pt, 1pt) circle (2pt);
				\filldraw[color=black, fill=skyblue] (3pt, 2pt) circle (2pt);
				\filldraw[color=black, fill=skyblue] (6pt, 4pt) circle (2pt);
				\filldraw[color=black, fill=skyblue] (14pt, 9pt) circle (2pt);
				\filldraw[color=black, fill=skyblue] (28pt, 21pt) circle (2pt);
				\filldraw[color=black, fill=skyblue] (140pt, 106pt) circle (2pt);
				\filldraw[color=black, fill=skyblue] (280pt, 224pt) circle (2pt);
				
				\matrix [below left, draw=none] at (current bounding box.north east) {
					\node [bigskybluenode, label=right:CoSky 'SQL query' with 3 attributes] {}; \\
				};
			\end{tikzpicture}
		}
		\caption{CoSky response time for Sql query with 3 attributes}\label{fig:temps_de_reponse_de_CoSky_en_sql_avec_3_dimensions}
	\end{figure}

	To compare CoSky implementations with 3 dimensions, we consider figure~\ref{fig:temps_de_reponse_de_CoSky_en_sql_avec_3_dimensions}. Evaluation are made for cardinalities up to $1$ billion tuples. In the worst case, response time is a little more than $3$ hours. 
	
	\section{Conclusion}
	
	In this article, we have presented new efficient Skyline ranking methods.
	
	The first is an improvement of the id-idp method using dominance hierarchy, faster than the original. An algorithmic implementation has been proposed.
	
	The second is the RankSky method, the adaptation of Google's well-known PageRank algorithm, using the IPL algorithm. It is regretfully not possible to embed this method in a relational DBMS, but an algorithmic implementation have been proposed.
	
	The third is the CoSky method, based on both TOPSIS approach, coming from multi-criteria analysis, and the similarity measure Salton cosine, coming from information retrieval. A SQL and an algorithmic implementations have, nevertheless, been proposed.
	
	DeepSky algorithm has been introduced in order to find top-$k$ ranked Skyline points, \emph{i.e.} with the highest scores, using multilevel Skyline principle associated to a skyline ranking solution such as CoSky, Ransky or dp-idp methods.
	
	Seeing their relevance and performance highlighted during experimental evaluations, we think the exposed methods could be the subject of future work and a future open source algorithmic platform.
	
	
	\begin{appendices}

	\end{appendices}
	
	
	\bibliography{biblio}


\begin{thebibliography}{19}
\ifx \bisbn   \undefined \def \bisbn  #1{ISBN #1}\fi
\ifx \binits  \undefined \def \binits#1{#1}\fi
\ifx \bauthor  \undefined \def \bauthor#1{#1}\fi
\ifx \batitle  \undefined \def \batitle#1{#1}\fi
\ifx \bjtitle  \undefined \def \bjtitle#1{#1}\fi
\ifx \bvolume  \undefined \def \bvolume#1{\textbf{#1}}\fi
\ifx \byear  \undefined \def \byear#1{#1}\fi
\ifx \bissue  \undefined \def \bissue#1{#1}\fi
\ifx \bfpage  \undefined \def \bfpage#1{#1}\fi
\ifx \blpage  \undefined \def \blpage #1{#1}\fi
\ifx \burl  \undefined \def \burl#1{\textsf{#1}}\fi
\ifx \doiurl  \undefined \def \doiurl#1{\url{https://doi.org/#1}}\fi
\ifx \betal  \undefined \def \betal{\textit{et al.}}\fi
\ifx \binstitute  \undefined \def \binstitute#1{#1}\fi
\ifx \binstitutionaled  \undefined \def \binstitutionaled#1{#1}\fi
\ifx \bctitle  \undefined \def \bctitle#1{#1}\fi
\ifx \beditor  \undefined \def \beditor#1{#1}\fi
\ifx \bpublisher  \undefined \def \bpublisher#1{#1}\fi
\ifx \bbtitle  \undefined \def \bbtitle#1{#1}\fi
\ifx \bedition  \undefined \def \bedition#1{#1}\fi
\ifx \bseriesno  \undefined \def \bseriesno#1{#1}\fi
\ifx \blocation  \undefined \def \blocation#1{#1}\fi
\ifx \bsertitle  \undefined \def \bsertitle#1{#1}\fi
\ifx \bsnm \undefined \def \bsnm#1{#1}\fi
\ifx \bsuffix \undefined \def \bsuffix#1{#1}\fi
\ifx \bparticle \undefined \def \bparticle#1{#1}\fi
\ifx \barticle \undefined \def \barticle#1{#1}\fi
\bibcommenthead
\ifx \bconfdate \undefined \def \bconfdate #1{#1}\fi
\ifx \botherref \undefined \def \botherref #1{#1}\fi
\ifx \url \undefined \def \url#1{\textsf{#1}}\fi
\ifx \bchapter \undefined \def \bchapter#1{#1}\fi
\ifx \bbook \undefined \def \bbook#1{#1}\fi
\ifx \bcomment \undefined \def \bcomment#1{#1}\fi
\ifx \oauthor \undefined \def \oauthor#1{#1}\fi
\ifx \citeauthoryear \undefined \def \citeauthoryear#1{#1}\fi
\ifx \endbibitem  \undefined \def \endbibitem {}\fi
\ifx \bconflocation  \undefined \def \bconflocation#1{#1}\fi
\ifx \arxivurl  \undefined \def \arxivurl#1{\textsf{#1}}\fi
\csname PreBibitemsHook\endcsname

\bibitem[\protect\citeauthoryear{Borzsony
  et~al.}{2001}]{borzsonySkylineOperator2001}
\begin{bchapter}
\bauthor{\bsnm{Borzsony}, \binits{S.}},
\bauthor{\bsnm{Kossmann}, \binits{D.}},
\bauthor{\bsnm{Stocker}, \binits{K.}}:
\bctitle{The {{Skyline}} operator}.
In: \bbtitle{Proceedings 17th {{International Conference}} on {{Data
  Engineering}}},
pp. \bfpage{421}--\blpage{430}.
\bpublisher{IEEE Comput. Soc},
\blocation{Heidelberg, Germany}
(\byear{2001}).
\doiurl{10.1109/ICDE.2001.914855}
\end{bchapter}
\endbibitem

\bibitem[\protect\citeauthoryear{Bentley
  et~al.}{1978}]{bentleyAverageNumberMaxima1978}
\begin{barticle}
\bauthor{\bsnm{Bentley}, \binits{J.L.}},
\bauthor{\bsnm{Kung}, \binits{H.T.}},
\bauthor{\bsnm{Schkolnick}, \binits{M.}},
\bauthor{\bsnm{Thompson}, \binits{C.D.}}:
\batitle{On the {{Average Number}} of {{Maxima}} in a {{Set}} of {{Vectors}}
  and {{Applications}}}.
\bjtitle{Journal of the ACM}
\bvolume{25}(\bissue{4}),
\bfpage{536}--\blpage{543}
(\byear{1978})
\doiurl{10.1145/322092.322095}
\end{barticle}
\endbibitem

\bibitem[\protect\citeauthoryear{Martin~Nevot and
  Lakhal}{2024}]{martinnevotClassementDobjetsSkylines2024}
\begin{bchapter}
\bauthor{\bsnm{Martin~Nevot}, \binits{M.}},
\bauthor{\bsnm{Lakhal}, \binits{L.}}:
\bctitle{Classement d'objets {{Skylines}} dans les bases de donn{\'e}es}.
In: \bbtitle{40{\`e}me Conf{\'e}rence Sur La {{Gestion}} de {{Donn{\'e}es}}},
\bconflocation{Orl{\'e}ans, France}
(\byear{2024})
\end{bchapter}
\endbibitem

\bibitem[\protect\citeauthoryear{Chan
  et~al.}{2006}]{chanHighDimensionalSkylines2006}
\begin{bchapter}
\bauthor{\bsnm{Chan}, \binits{C.-Y.}},
\bauthor{\bsnm{Jagadish}, \binits{H.V.}},
\bauthor{\bsnm{Tan}, \binits{K.-L.}},
\bauthor{\bsnm{Tung}, \binits{A.K.H.}},
\bauthor{\bsnm{Zhang}, \binits{Z.}}:
\bctitle{On {{High Dimensional Skylines}}}.
In: \beditor{\bsnm{Hutchison}, \binits{D.}},
\beditor{\bsnm{Kanade}, \binits{T.}},
\beditor{\bsnm{Kittler}, \binits{J.}},
\beditor{\bsnm{Kleinberg}, \binits{J.M.}},
\beditor{\bsnm{Mattern}, \binits{F.}},
\beditor{\bsnm{Mitchell}, \binits{J.C.}},
\beditor{\bsnm{Naor}, \binits{M.}},
\beditor{\bsnm{Nierstrasz}, \binits{O.}},
\beditor{\bsnm{Pandu~Rangan}, \binits{C.}},
\beditor{\bsnm{Steffen}, \binits{B.}},
\beditor{\bsnm{Sudan}, \binits{M.}},
\beditor{\bsnm{Terzopoulos}, \binits{D.}},
\beditor{\bsnm{Tygar}, \binits{D.}},
\beditor{\bsnm{Vardi}, \binits{M.Y.}},
\beditor{\bsnm{Weikum}, \binits{G.}},
\beditor{\bsnm{Ioannidis}, \binits{Y.}},
\beditor{\bsnm{Scholl}, \binits{M.H.}},
\beditor{\bsnm{Schmidt}, \binits{J.W.}},
\beditor{\bsnm{Matthes}, \binits{F.}},
\beditor{\bsnm{Hatzopoulos}, \binits{M.}},
\beditor{\bsnm{Boehm}, \binits{K.}},
\beditor{\bsnm{Kemper}, \binits{A.}},
\beditor{\bsnm{Grust}, \binits{T.}},
\beditor{\bsnm{Boehm}, \binits{C.}} (eds.)
\bbtitle{Advances in {{Database Technology}} - {{EDBT}} 2006},
vol. \bseriesno{3896},
pp. \bfpage{478}--\blpage{495}.
\bpublisher{Springer},
\blocation{Berlin, Heidelberg}
(\byear{2006}).
\doiurl{10.1007/11687238_30}
\end{bchapter}
\endbibitem

\bibitem[\protect\citeauthoryear{Yiu and
  Mamoulis}{2007}]{yiuEfficientProcessingTopk2007}
\begin{bchapter}
\bauthor{\bsnm{Yiu}, \binits{M.}},
\bauthor{\bsnm{Mamoulis}, \binits{N.}}:
\bctitle{Efficient processing of top-k dominating queries on multi-dimensional
  data}.
In: \bbtitle{Proceedings of the 33rd {{International Conference}} on {{Very
  Large Data Bases}}, {{University}} of {{Vienna}}, {{Austria}}, {{September}}
  23-27, 2007},
pp. \bfpage{483}--\blpage{494}.
\bpublisher{ACM},
\blocation{Vienna, Austria}
(\byear{2007})
\end{bchapter}
\endbibitem

\bibitem[\protect\citeauthoryear{Bartolini
  et~al.}{2007}]{bartoliniFlexibleIntegrationMultimedia2007}
\begin{barticle}
\bauthor{\bsnm{Bartolini}, \binits{I.}},
\bauthor{\bsnm{Ciaccia}, \binits{P.}},
\bauthor{\bsnm{Oria}, \binits{V.}},
\bauthor{\bsnm{{\"O}zsu}, \binits{M.T.}}:
\batitle{Flexible integration of multimedia sub-queries with qualitative
  preferences}.
\bjtitle{Multimedia Tools and Applications}
\bvolume{33}(\bissue{3}),
\bfpage{275}--\blpage{300}
(\byear{2007})
\doiurl{10.1007/s11042-007-0103-1}
\end{barticle}
\endbibitem

\bibitem[\protect\citeauthoryear{Lakhal
  et~al.}{2017}]{lakhalMultidimensionalSkylineAnalysis2017}
\begin{barticle}
\bauthor{\bsnm{Lakhal}, \binits{L.}},
\bauthor{\bsnm{Nedjar}, \binits{S.}},
\bauthor{\bsnm{Cicchetti}, \binits{R.}}:
\batitle{Multidimensional skyline analysis based on agree concept lattices}.
\bjtitle{Intelligent Data Analysis}
\bvolume{21}(\bissue{5}),
\bfpage{1245}--\blpage{1265}
(\byear{2017})
\doiurl{10.3233/IDA-163111}
\end{barticle}
\endbibitem

\bibitem[\protect\citeauthoryear{Vlachou and
  Vazirgiannis}{2010}]{vlachouRankingSkyDiscovering2010}
\begin{barticle}
\bauthor{\bsnm{Vlachou}, \binits{A.}},
\bauthor{\bsnm{Vazirgiannis}, \binits{M.}}:
\batitle{Ranking the sky: {{Discovering}} the importance of skyline points
  through subspace dominance relationships}.
\bjtitle{Data \& Knowledge Engineering}
\bvolume{69}(\bissue{9}),
\bfpage{943}--\blpage{964}
(\byear{2010})
\doiurl{10.1016/j.datak.2010.03.008}
\end{barticle}
\endbibitem

\bibitem[\protect\citeauthoryear{Gao
  et~al.}{2015}]{gaoEfficientAlgorithmsFinding2015}
\begin{barticle}
\bauthor{\bsnm{Gao}, \binits{Y.}},
\bauthor{\bsnm{Liu}, \binits{Q.}},
\bauthor{\bsnm{Chen}, \binits{L.}},
\bauthor{\bsnm{Chen}, \binits{G.}},
\bauthor{\bsnm{Li}, \binits{Q.}}:
\batitle{Efficient algorithms for finding the most desirable skyline objects}.
\bjtitle{Knowledge-Based Systems}
\bvolume{89},
\bfpage{250}--\blpage{264}
(\byear{2015})
\doiurl{10.1016/j.knosys.2015.07.007}
\end{barticle}
\endbibitem

\bibitem[\protect\citeauthoryear{Valkanas
  et~al.}{2014}]{valkanasSkylineRankingIR2014}
\begin{bchapter}
\bauthor{\bsnm{Valkanas}, \binits{G.}},
\bauthor{\bsnm{Papadopoulos}, \binits{A.N.}},
\bauthor{\bsnm{Gunopulos}, \binits{D.}}:
\bctitle{Skyline ranking {\`a} la {{IR}}}.
In: \beditor{\bsnm{Candan}, \binits{K.S.}},
\beditor{\bsnm{{Amer-Yahia}}, \binits{S.}},
\beditor{\bsnm{Schweikardt}, \binits{N.}},
\beditor{\bsnm{Christophides}, \binits{V.}},
\beditor{\bsnm{Leroy}, \binits{V.}} (eds.)
\bbtitle{Proceedings of the Workshops of the {{EDBT}}/{{ICDT}} 2014 Joint
  Conference ({{EDBT}}/{{ICDT}} 2014), Athens, Greece, March 28, 2014}.
\bsertitle{{{CEUR}} Workshop Proceedings},
vol. \bseriesno{1133},
pp. \bfpage{182}--\blpage{187}.
\bpublisher{CEUR-WS.org}, \blocation{???}
(\byear{2014})
\end{bchapter}
\endbibitem

\bibitem[\protect\citeauthoryear{Page
  et~al.}{1998}]{pagePageRankCitationRanking1998}
\begin{botherref}
\oauthor{\bsnm{Page}, \binits{L.}},
\oauthor{\bsnm{Brin}, \binits{S.}},
\oauthor{\bsnm{Motwani}, \binits{R.}},
\oauthor{\bsnm{Winograd}, \binits{T.}}:
The {{PageRank Citation Ranking}}: {{Bringing Order}} to the {{Web}}.
Technical report,
Stanford Digital Libraries Technologies Project
(1998)
\end{botherref}
\endbibitem

\bibitem[\protect\citeauthoryear{Langville and
  Meyer}{2006}]{langvilleGooglesPageRankScience2006}
\begin{bbook}
\bauthor{\bsnm{Langville}, \binits{A.N.}},
\bauthor{\bsnm{Meyer}, \binits{C.D.}}:
\bbtitle{Google's {{PageRank}} and Beyond: The Science of Search Engine
  Rankings}.
\bpublisher{Princeton University Press},
\blocation{Princeton, N.J}
(\byear{2006})
\end{bbook}
\endbibitem

\bibitem[\protect\citeauthoryear{Lai et~al.}{1994}]{laiTOPSISMODM1994}
\begin{barticle}
\bauthor{\bsnm{Lai}, \binits{Y.-J.}},
\bauthor{\bsnm{Liu}, \binits{T.-Y.}},
\bauthor{\bsnm{Hwang}, \binits{C.-L.}}:
\batitle{{{TOPSIS}} for {{MODM}}}.
\bjtitle{European Journal of Operational Research}
\bvolume{76}(\bissue{3}),
\bfpage{486}--\blpage{500}
(\byear{1994})
\doiurl{10.1016/0377-2217(94)90282-8}
\end{barticle}
\endbibitem

\bibitem[\protect\citeauthoryear{Behzadian
  et~al.}{2012}]{behzadianStateofTheartSurvey2012}
\begin{barticle}
\bauthor{\bsnm{Behzadian}, \binits{M.}},
\bauthor{\bsnm{Khanmohammadi~Otaghsara}, \binits{S.}},
\bauthor{\bsnm{Yazdani}, \binits{M.}},
\bauthor{\bsnm{Ignatius}, \binits{J.}}:
\batitle{A state-of the-art survey of {{TOPSIS}} applications}.
\bjtitle{Expert Systems with Applications}
\bvolume{39}(\bissue{17}),
\bfpage{13051}--\blpage{13069}
(\byear{2012})
\doiurl{10.1016/j.eswa.2012.05.056}
\end{barticle}
\endbibitem

\bibitem[\protect\citeauthoryear{Huang}{2008}]{huangCombiningEntropyWeight2008}
\begin{bchapter}
\bauthor{\bsnm{Huang}, \binits{J.}}:
\bctitle{Combining entropy weight and {{TOPSIS}} method for information system
  selection}.
In: \bbtitle{2008 {{IEEE Conference}} on {{Cybernetics}} and {{Intelligent
  Systems}}},
pp. \bfpage{1281}--\blpage{1284}.
\bpublisher{IEEE},
\blocation{Chengdu, China}
(\byear{2008}).
\doiurl{10.1109/ICCIS.2008.4670971}
\end{bchapter}
\endbibitem

\bibitem[\protect\citeauthoryear{Lotfi and
  Fallahnejad}{2010}]{lotfiImpreciseShannonsEntropy2010}
\begin{barticle}
\bauthor{\bsnm{Lotfi}, \binits{F.H.}},
\bauthor{\bsnm{Fallahnejad}, \binits{R.}}:
\batitle{Imprecise {{Shannon}}'s {{Entropy}} and {{Multi Attribute Decision
  Making}}}.
\bjtitle{Entropy}
\bvolume{12}(\bissue{1}),
\bfpage{53}--\blpage{62}
(\byear{2010})
\doiurl{10.3390/e12010053}
\end{barticle}
\endbibitem

\bibitem[\protect\citeauthoryear{Papadias
  et~al.}{2005}]{papadiasProgressiveSkylineComputation2005}
\begin{barticle}
\bauthor{\bsnm{Papadias}, \binits{D.}},
\bauthor{\bsnm{Tao}, \binits{Y.}},
\bauthor{\bsnm{Fu}, \binits{G.}},
\bauthor{\bsnm{Seeger}, \binits{B.}}:
\batitle{Progressive skyline computation in database systems}.
\bjtitle{ACM Transactions on Database Systems}
\bvolume{30}(\bissue{1}),
\bfpage{41}--\blpage{82}
(\byear{2005})
\doiurl{10.1145/1061318.1061320}
\end{barticle}
\endbibitem

\bibitem[\protect\citeauthoryear{Preisinger and
  Endres}{2015}]{preisingerLookingBestNot2015}
\begin{botherref}
\oauthor{\bsnm{Preisinger}, \binits{T.}},
\oauthor{\bsnm{Endres}, \binits{M.}}:
Looking for the best, but not too many of them: Multi-level and top-k skylines.
International Journal on Advances in Software,
467--480
(2015)
\end{botherref}
\endbibitem

\bibitem[\protect\citeauthoryear{Fabris}{2022}]{fabrisFlexibleSkylinesRegret2022}
\begin{botherref}
\oauthor{\bsnm{Fabris}, \binits{V.}}:
Flexible Skylines, Regret Minimization and Skyline Ranking: A Comparison to
  Know How to Select the Right Approach.
arXiv
(2022).
\doiurl{10.48550/ARXIV.2201.10179}
\end{botherref}
\endbibitem

\end{thebibliography}
	
\end{document}